\documentclass[12pt,english,floatfix,nofootinbib,superscriptaddress,aps,prd,preprint]{revtex4}
\usepackage[utf8]{inputenc}
\usepackage{float}
\usepackage{array}
\usepackage{lipsum}
\usepackage{graphicx}
\usepackage{amsmath,amsthm,amsfonts,amssymb}
\usepackage{graphicx}
\usepackage[english]{babel}
\usepackage{color}
\usepackage{tensor}
\usepackage{esint}
\usepackage[dvips]{epsfig}
\usepackage[dvips]{graphicx}
\usepackage{float}
\usepackage{units}
\usepackage{textcomp}
\usepackage{mathrsfs}
\usepackage{amsmath}
\usepackage[makeroom]{cancel}
\usepackage{amssymb}
\usepackage{amsbsy}
\usepackage{amsfonts}
\usepackage{amssymb,mathrsfs,xcolor}
\usepackage{esint}
\usepackage{array}
\usepackage{graphicx}

\usepackage{hyperref}
\hypersetup{
    colorlinks,
    citecolor=blue,
    filecolor=green,
    linkcolor=purple,
    urlcolor=red,
}

\usepackage{slashed}

\newcommand{\ie}{\begin{equation}}
\newcommand{\fe}{\end{equation}}
\newcommand{\se}{\begin{eqnarray}}
\newcommand{\ff}{\end{eqnarray}}

\usepackage{hyperref}
\hypersetup{colorlinks,breaklinks,
			citecolor=[rgb]{0,0.0,1.0},
            urlcolor=[rgb]{0.0,0.0,1.0},
            linkcolor=[rgb]{0,0.5,0.9}}

\begin{document}

\title{Bouncing universe in a heat bath}

\author{A. A. Ara\'{u}jo Filho}
\email{dilto@fisica.ufc.br}

\affiliation{Universidade Federal do Cear\'a (UFC), Departamento de F\'isica,\\ Campus do Pici,
Fortaleza - CE, C.P. 6030, 60455-760 - Brazil.}

\author{A. Yu. Petrov}
\email{petrov@fisica.ufpb.br}
\affiliation{Departamento de Física, Universidade Federal da Paraíba, Caixa Postal 5008, 58051-970, João Pessoa, Paraíba,  Brazil.}

%%%%%%%%%%%%%%%%%%%%%%%%%%%%%%%%%%%%%%%%%%%%%%%%%%%%%%%%%%%%%%%%%%%%%%

%%%%%%%%%%%%%%%%%%%%%%%%%%%%%%%%%%%%%%%%%%%%%%%%%%%%%%%%%%%%%%%%%%%%%%

%%%%%%%%%%%%%%%%%%%%%%%%%%%%%%%%%%%%%%%%%%%%%%%%%%%%%%%%%%%%%%%%%%%%%%%%%%%%%%%%%%%%%%%%%%%%%%%%%%%%%%%%%%%%%%%%%%%%%%%%%%%%%%%%%%%%%%%%%%%%%%%%%%%%%%%%%%%%%%%%%%%%%%%%%%%%%%%%%%%%%%%%%%%%%%%%%%%%%%%%%%%%
%%%%%%%%%%%%%%%%%%%%%%%%%%%%%%%%%%%%%%%%%%%%%%%%%%%

\date{\today}

\begin{abstract}

We develop a thermal description for photon modes within the context of bouncing universe. Within this study, we start with a Lorentz-breaking dispersion relation which accounts for modified Friedmann equations with a bounce solution. We calculate the spectral radiance, the entropy, the Helmholtz free energy, the heat capacity, and the mean energy. Nevertheless, the latter two ones turn out to contribute only in a trivial manner. Furthermore, one intriguing aspect gives rise to: for some configurations of $\eta$, $l_{P}$, and $T$, the thermodynamical behavior of the system seems to indicate instability. However, despite of showing this particularity, the solutions of the thermodynamic functions in general turn out to agree with the literature, e.g., the second law of thermodynamics is maintained. More so, all the results are derived \textit{analytically} and three different regimes of temperature of the universe are also considered, i.e., the inflationary era, the electroweak epoch and the cosmic microwave background.

\end{abstract}

%%\keywords{Thermodynamics properties; Higher-Derivative, Poldolsky eletrodynamics.}

\maketitle

%%%%%%%%%%%%%%%%%%%%%%%%%%%%%%%%%%%%%%%%%%%%%%%%%%%%%%%%%%%%%%%%%%%%%%%%%%%%%%%%%%%%%%%%%%%%%%%%%%%%%%%%%%%%%%%%%%%%%%%%%%%%%%%%%%%%%%%%%%%%%%%%%%%%%%%%%%%%%%%%%%%%%%%%%%%%%%%%%%%%%%%%%%%%%%%%%%%%%%%%%%%%%%%%%%%%%%%%%%%%%%%%%%%%%%%%%%%%%%%%%%%%%%%%%%%%%%%%%%%%%%%%%%%%%%%%%%%%%%%%%%%%%%%%%%%%%%%%%%%%%%%%%%%%%%%%%%%%%%%%%%%%%%%%%%%%%%%%%%%%%%%%%%%%%%%%%%%%%%%%%%%%%%%%%%%%%%%%%%%%%%%%%%%%%%%%%%%%%%%%%%%%%%%%%%%%%%%%%%%%%%%%%%%%%%%%%%%%%%%%%%%%%%%%%%%%%%%%%%%%%%%%%%%%%%%%%%%%%%%%%%%%%%

\section{Introduction}

The idea that the widely-spread gravitational theory may be outlined by entropy, was formulated for the first time by Jacobson \cite{jacobson1995thermodynamics} and afterwards received further developments \cite{ling2009bouncing,cai2005first,eling2006nonequilibrium,padmanabhan2002classical,akbar2007thermodynamic,sheykhi2007deep,ge2007first,guedens2012horizon,cai2008corrected,sharif2012thermodynamics}. Additionally, in order to overcome the longstanding issue of the cosmological singularity present within the Big Bounce model, a notable approach was used considering the semiclassical limit \cite{bojowald2008quantum,ashtekar2006quantum,ashtekar2006quantum2,ashtekar2006quantum3}. 

For instance, in Ref. \cite{ling2009bouncing}, the authors assumed, as the starting point, a certain dispersion relation rather than a particular Lagrangian. With this, a remarkable aspect emerged: the avoidance of divergences in cosmological scenarios when the high energy limit was taken into account. In other words, they obtained a bouncing solution, i.e., with the absence of singularity, ascribed to the modification of the Friedmann equations. This procedure is absolutely reasonable since, being in agreement with previous studies, starting only from such modified dispersion relations can be considered though as an alternative way to investigate for instance cosmological scenarios \cite{amelino2006black, aa2020lorentz, amelino2001testable}.

Moreover, one significant aspect which is worth investigating in the context of Lorentz symmetry violation is actually the respective thermal aspects of the corresponding theory. With such study, we can obtain a better comprehension of how massless and massive modes behave when different range of temperatures are taken into account. This might possibly give us additional information to confront the theoretical background with the experimental results in order to help searching for any trace of the Lorentz symmetry violation. Furthermore, the investigation based on the thermodynamic properties and the Lorentz-violating (LV) effects could supply further knowledge concerning primordial stages of expansion of the universe, whose size is consistent with characteristic scales of Lorentz symmetry breaking \cite{kostelecky2011data}.

Initially, the procedure to carry out the thermodynamic aspects involving Lorentz violation regarding statistical mechanics was proposed in Ref. \cite{colladay2004statistical}. After that, several works using such method have been developed within different scenarios \cite{casana2008lorentz,casana2009finite,gomes2010free,araujo2021thermal, araujo2021thermodynamic,aa2020lorentz,anacleto2018lorentz,das2009relativistic,petrov2021higher}. Nevertheless, in the context of Ref. \cite{colladay2004statistical}, the examination of the thermodynamic properties concerning bouncing universe cases has not yet been explored up to date.
In this paper, we study the thermal behavior of a photon gas in a bouncing universe, for a certain form of dispersion relation.

The structure of this paper looks like follows. In the section \ref{secII}, we define our model using the patterns of doubly special relativity, and perform calculations of the thermodynamical quantities; finally, in the section \ref{conclusion}, we highlight the main results developed in the manuscript.

\section{Modified dispersion relations and their thermodynamical impacts} \label{secII}

\begin{figure}[tbh]
\centering
\includegraphics[width=8cm,height=5cm]{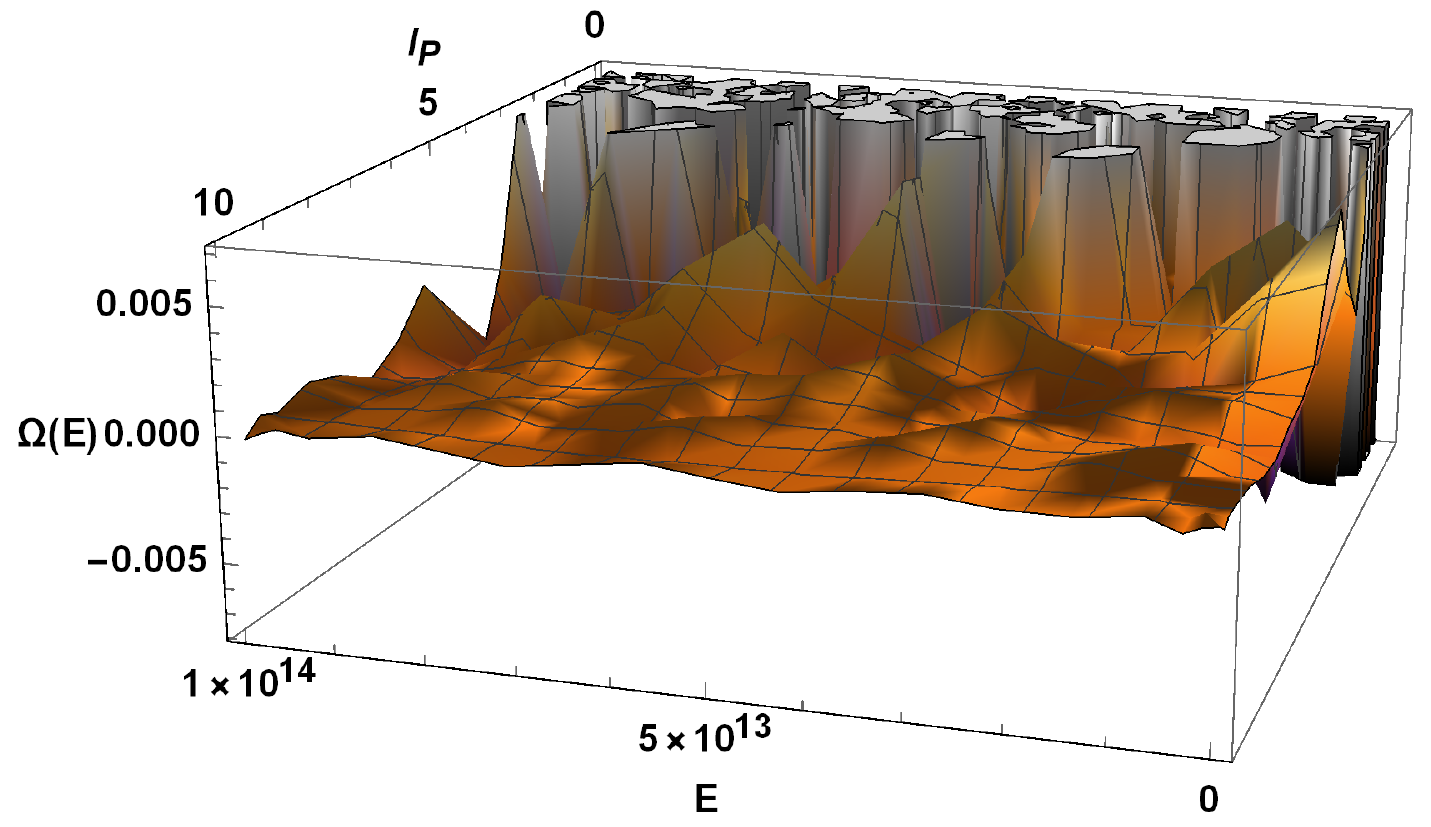}
\includegraphics[width=8cm,height=5cm]{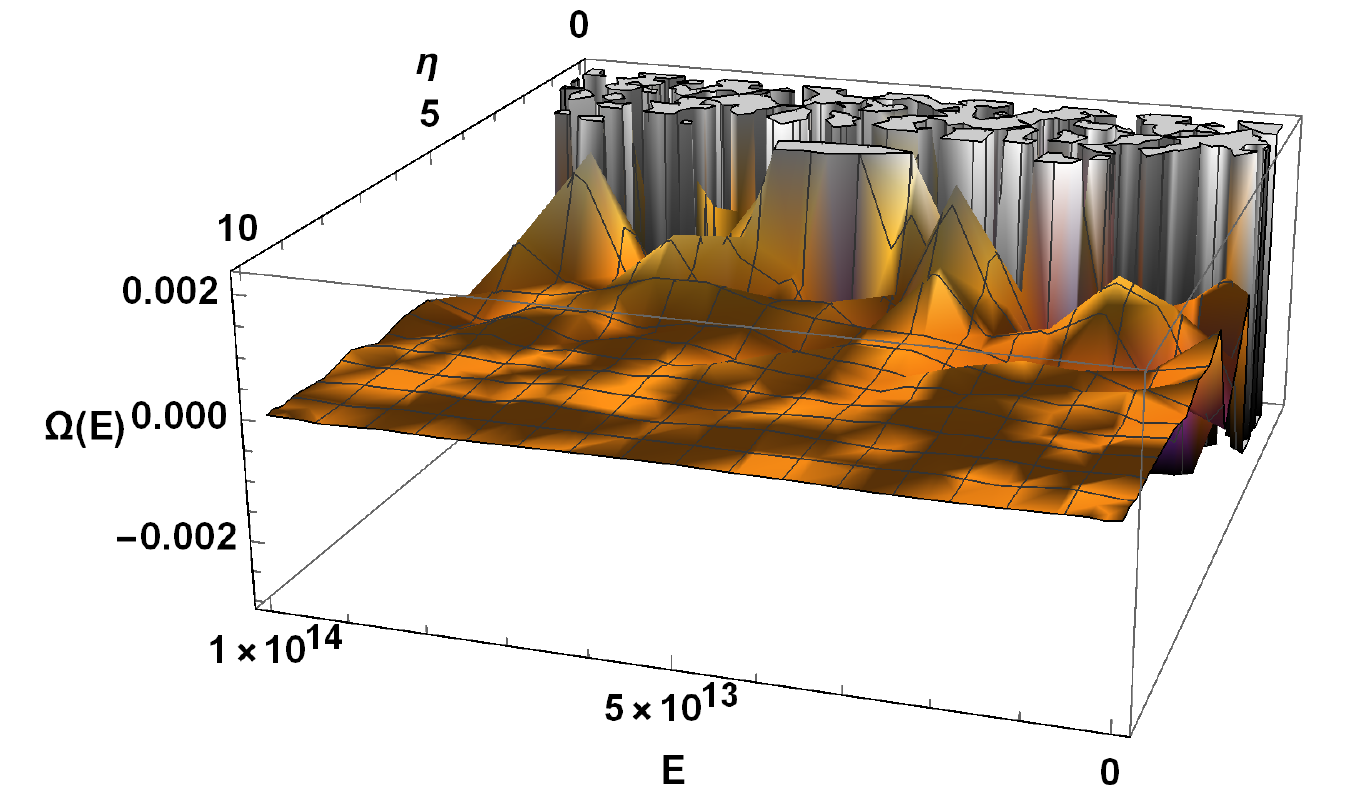}
\caption{The accessible states of the system to the inflationary period.}
\label{acessiblestates-inflation}
\end{figure}

\begin{figure}[tbh]
\centering
\includegraphics[width=8cm,height=5cm]{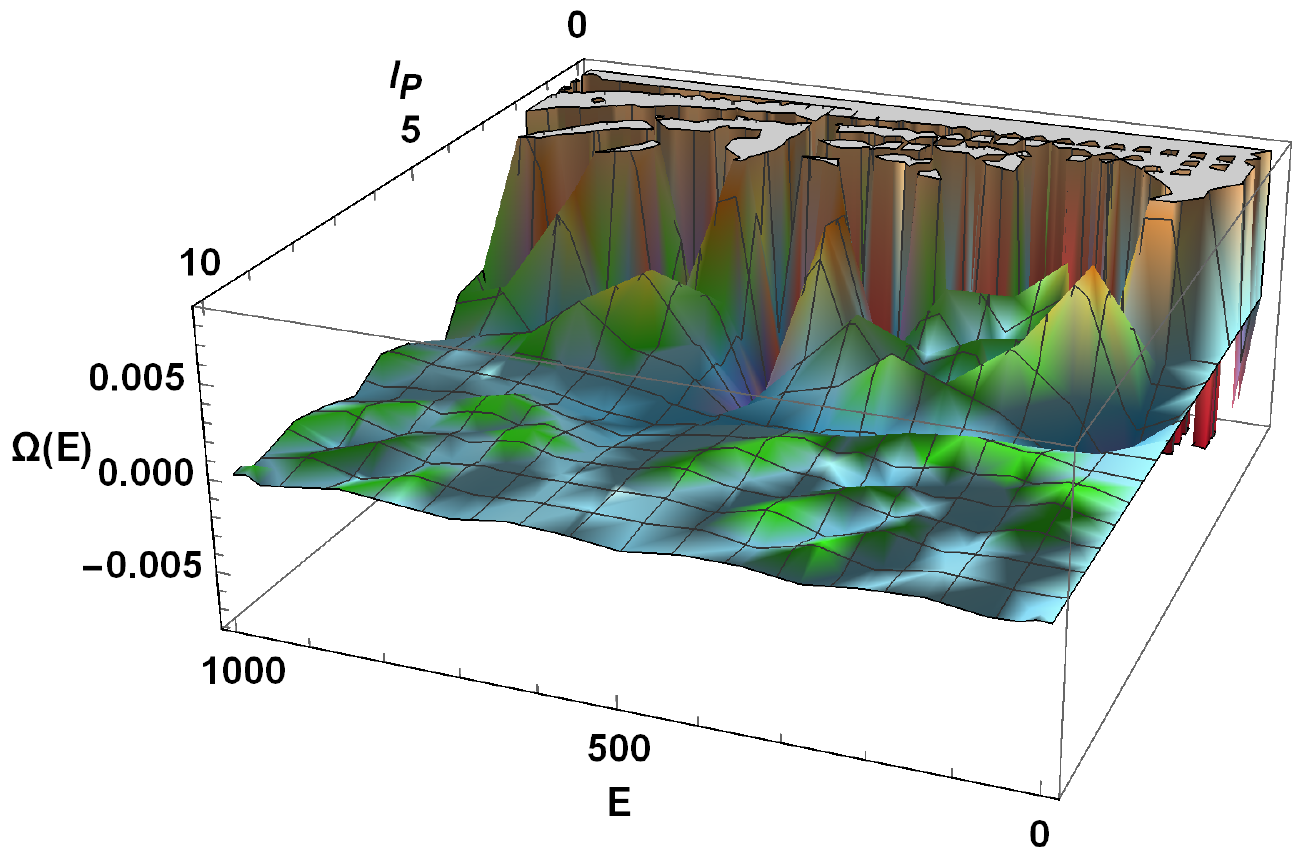}
\includegraphics[width=8cm,height=5cm]{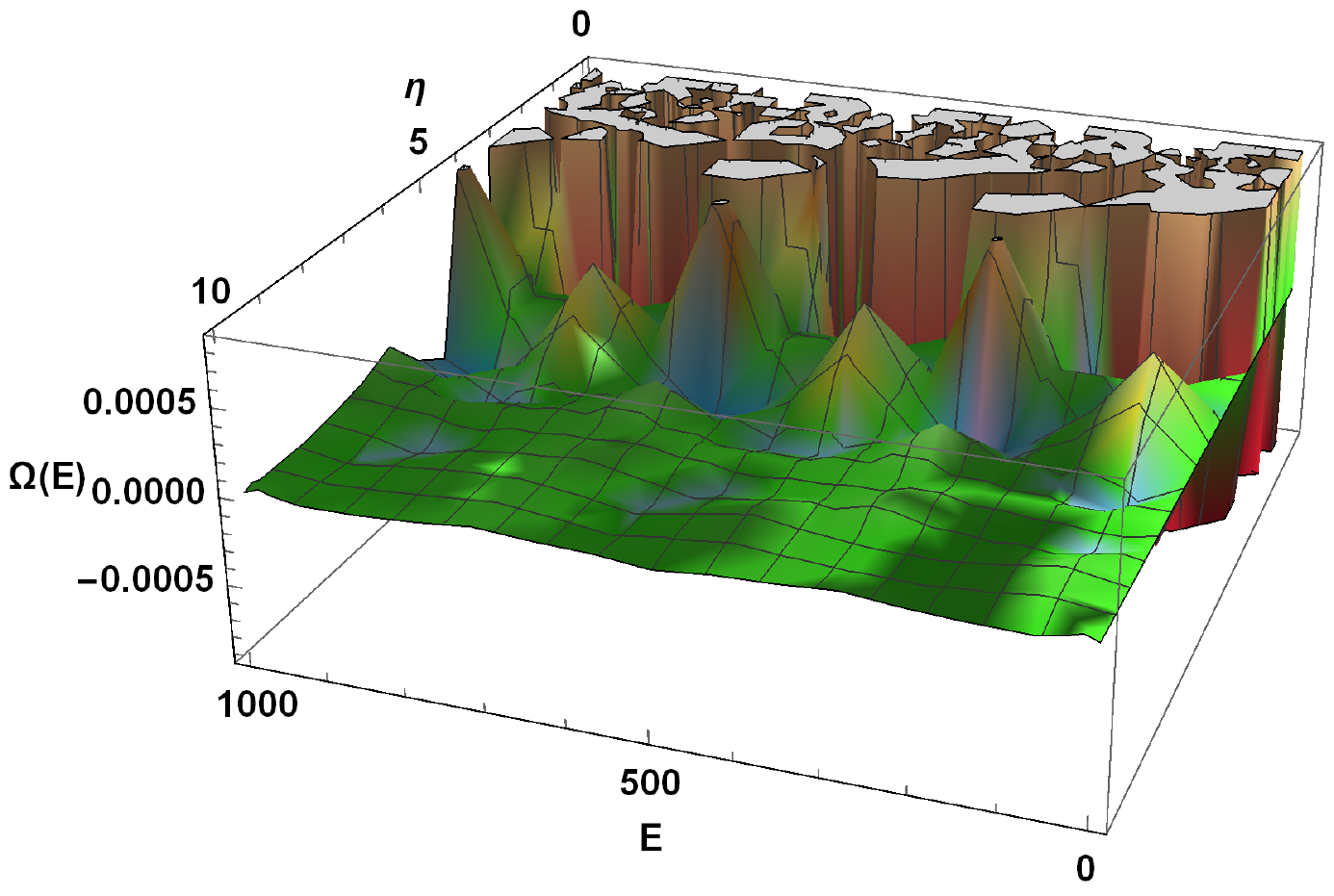}
\caption{The accessible states of the system to the electroweak period.}
\label{acessiblestates-electroweak}
\end{figure}

\begin{figure}[tbh]
\centering
\includegraphics[width=8cm,height=5cm]{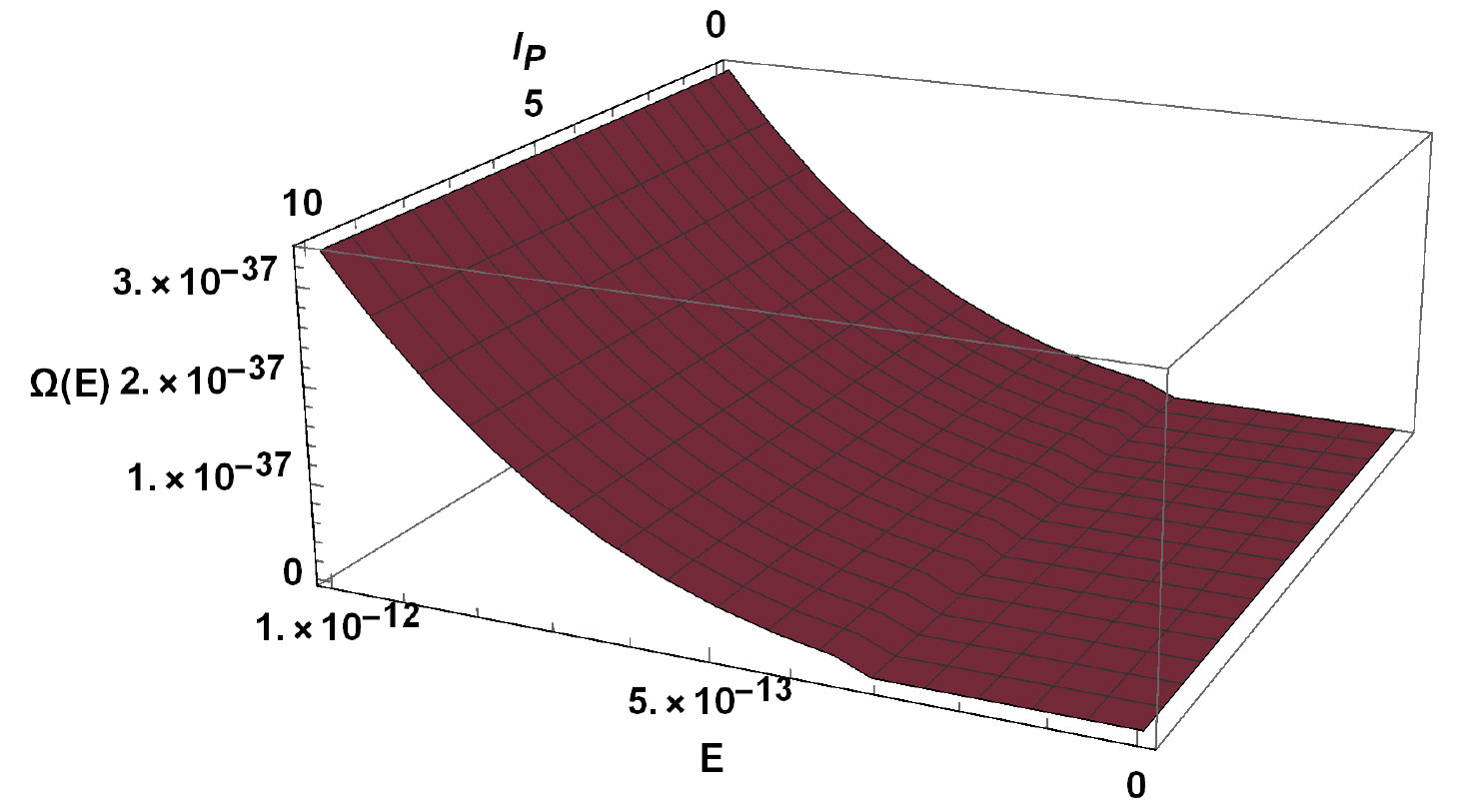}
\includegraphics[width=8cm,height=5cm]{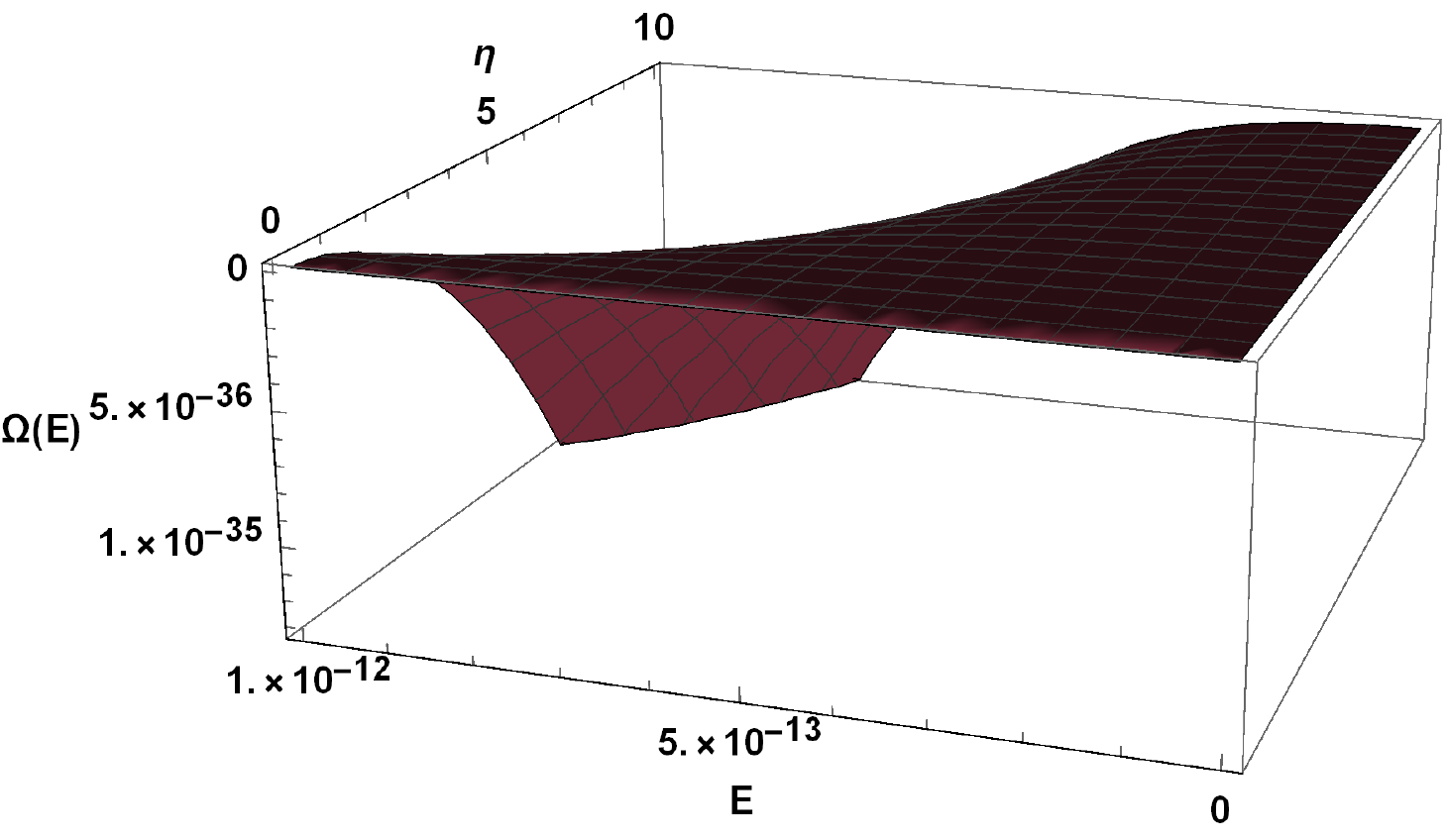}
\caption{The accessible states of the system to the cosmic microwave background.}
\label{acessiblestates-cmb}
\end{figure}

Here, we start from a general modified dispersion relation in the context of \textit{doubly special relativity} (DSR) \cite{amelino2002relativity,magueijo2004gravity,pan2016bouncing,feng2020thermodynamics}
\ie
E^{2} f^{2}(l_{P}E)=  k^{2} g^{2}(l_{P}E) + m^{2}
\label{modifydisrel}
\fe
where $E$ is the energy, $l_{P}$ is the Planck length, defining the fundamental energy scale playing the key role within the DSR, and $m$ is the mass; more so, $f(l_{P}E)$, $g(l_{P}E)$ are arbitrary functions and $k$ is the momentum of the particle. 
The key idea of this concept (for motivations and more detailed discussions, see f.e. \cite{amelino2002relativity}) is that the dispersion relations are assumed to be characterized by an additional constant, describing the characteristic energy scale -- the Planck energy. As a result, the dispersion relations, at small energies, behave as usual ones, while at higher energies, their deformation, and consequent difference from the usual relativistic scenario, becomes to be the crucial effect. Due to the presence of two constants, that is, the speed of light and the energy scale, this concept was defined as doubly special relativity.
It is worth mentioning that the modified dispersion relation coming from Eq. (\ref{modifydisrel}) has been used as a basis for investigations of black holes \cite{amelino2006black} and rainbow spacetimes \cite{ling2006modified,han2008modified,ling2007rainbow}. Note that, if one considers the limit where $f(l_{P}E) = g(l_{P}E) \approx 1$, i.e., within a low energy domain $1 \gg l_{P}E$, one naturally recovers the usual massive dispersion relation $E^{2} - k^{2} = m^{2}$. An example of a specific choice of the parameters $f(l_{P}E)$ and $g(l_{P}E)$, was done in Ref. \cite{ling2009bouncing,zhang2018quantum}, explicitly,
\ie
g(l_{P}E) = 1,    \,\,\,\,\,\, f(l_{P}E) = \sqrt{\frac{\sin^{2} (\eta^{2} l_{P}E)}{\eta^{2} l^{2}_{P}E^{2}}} \label{sin2}
\fe
which implies
\ie
\frac{1}{\eta^{2} l_{P}} \sin(\eta^{2} l_{P} E) = \sqrt{k^{2} + m^{2}} \label{sin}
\fe
where $\eta$ is a dimensionless parameter. The specific choices of the functions $g(l_{P}E)$, and $f(l_{P}E)$ presented in Eq. (\ref{sin2}), as well as in the modified dispersion relation (\ref{sin}) are mainly motivated by the study of black hole thermodynamics \cite{han2008modified,amelino2004severe,ling2007thermodynamics}. Furthermore, as argued in \cite{ling2009bouncing}, such a choice of the modified dispersion relation {\bf can} generate bouncing solutions through a thermodynamical approach of general relativity with an apparent horizon. With this, the accessible state of the system, namely, $\Omega(E)$, can properly derived as 
\ie
\Omega(E) = \frac{\Gamma}{\pi^{2}} \int \frac{1}{\eta^{2} l^{2}_{P}} \sin^{2}(\eta^{2} l_{P} E)\cos(\eta^{2} l_{P} E)\, \mathrm{d}E \label{accessiblestates}
\fe   
where $\Gamma$ is the volume of the reservoir\footnote{For the sake of simplicity, hereafter, we will consider the following calculations in a per volume approach.}. Here, we are able to proceed with calculations derived from Eq. (\ref{accessiblestates}) in a \textit{analytical} manner. It is worth mentioning that such \textit{analytical} studies involving thermodynamic properties in the context of Lorentz symmetry violation were performed only in rare cases. Moreover, from Figs. \ref{acessiblestates-inflation}, \ref{acessiblestates-electroweak}, \ref{acessiblestates-cmb}, we accomplish an analysis taking into account low through high-energy temperatures to the accessible states of the system, e.g., we consider three distinct periods of the universe: the inflationary period ($T = 10^{13}$ GeV), the electroweak epoch ($T = 10^{3}$ GeV), and cosmic microwave background radiation ($T = 10^{-13}$ GeV).

\begin{figure}[tbh]
\centering
\includegraphics[width=8cm,height=5cm]{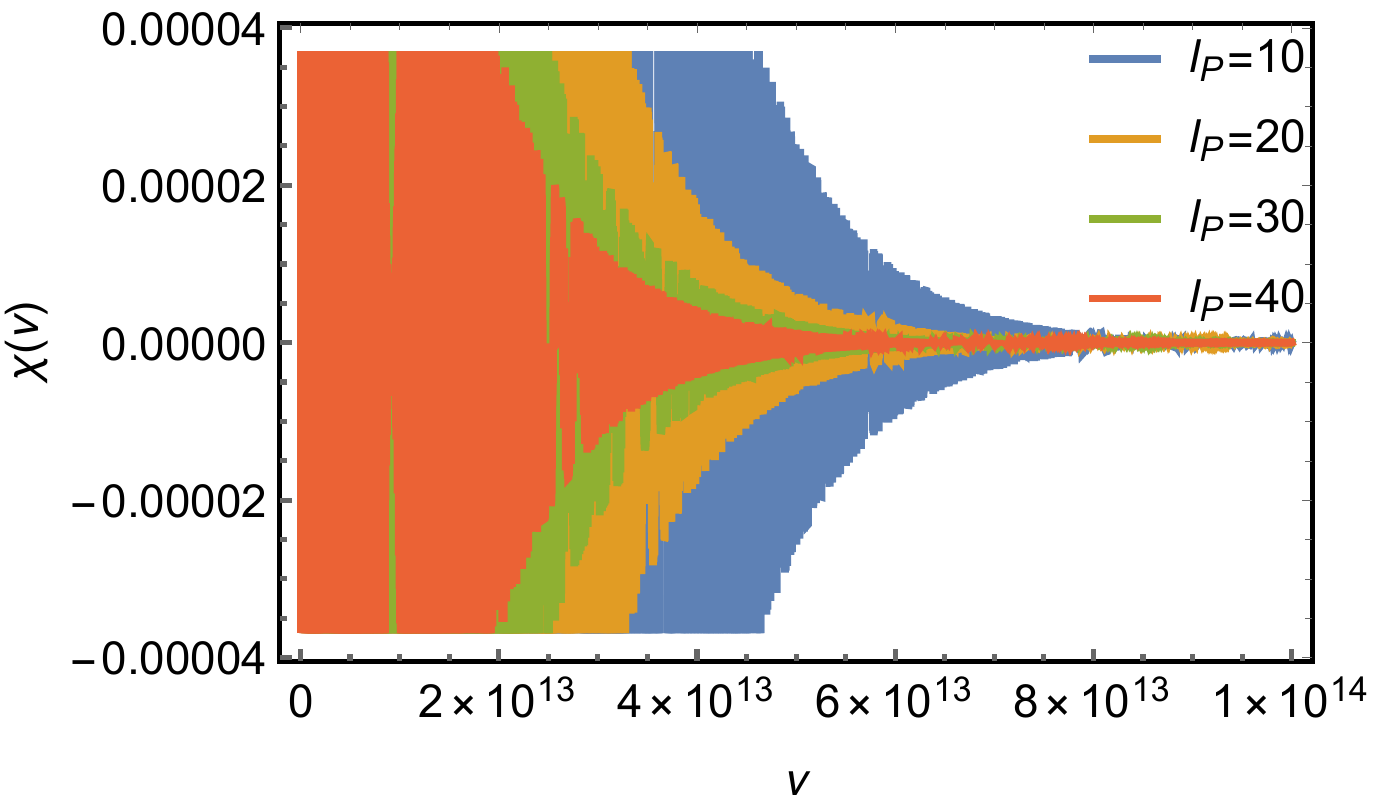}
\includegraphics[width=8cm,height=5cm]{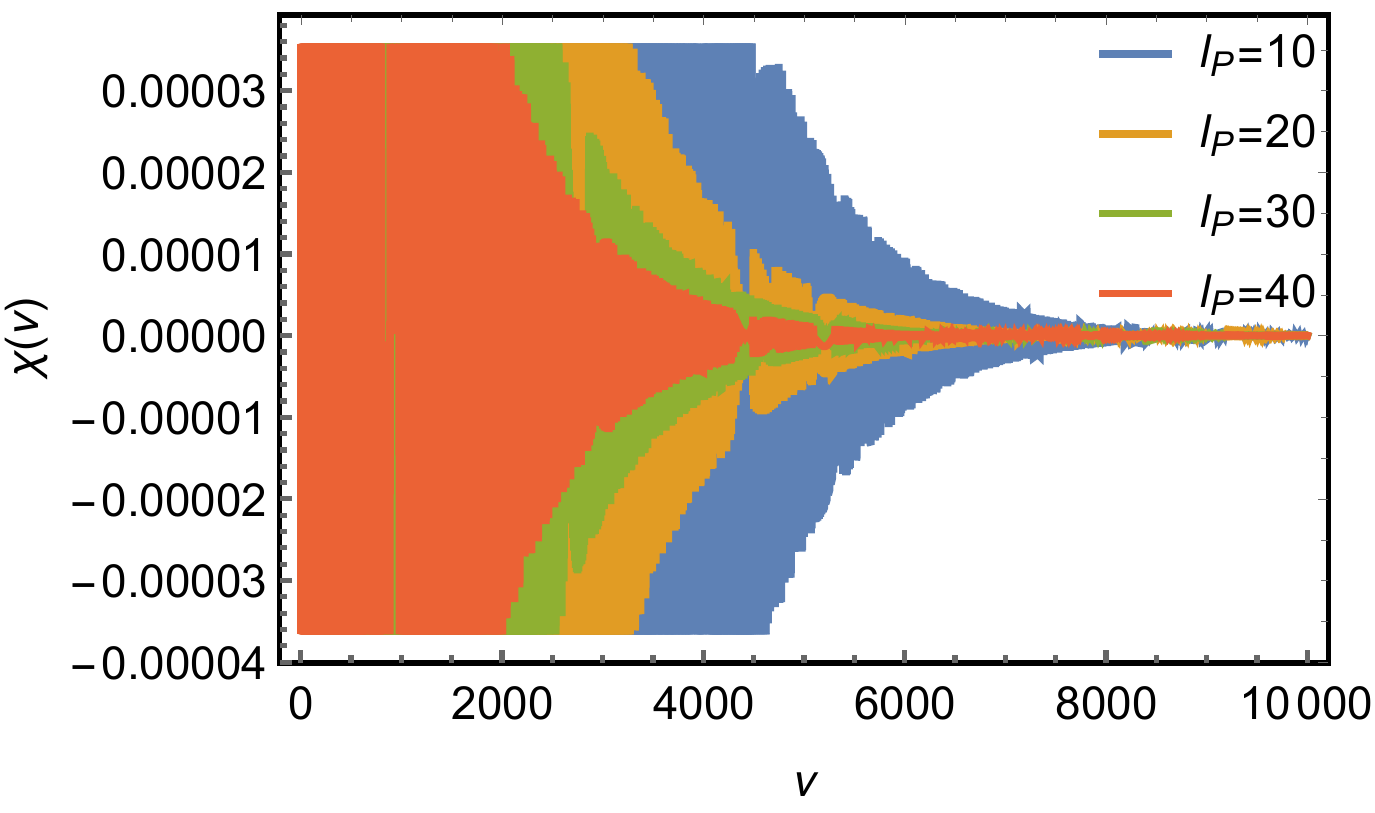}
\includegraphics[width=8cm,height=5cm]{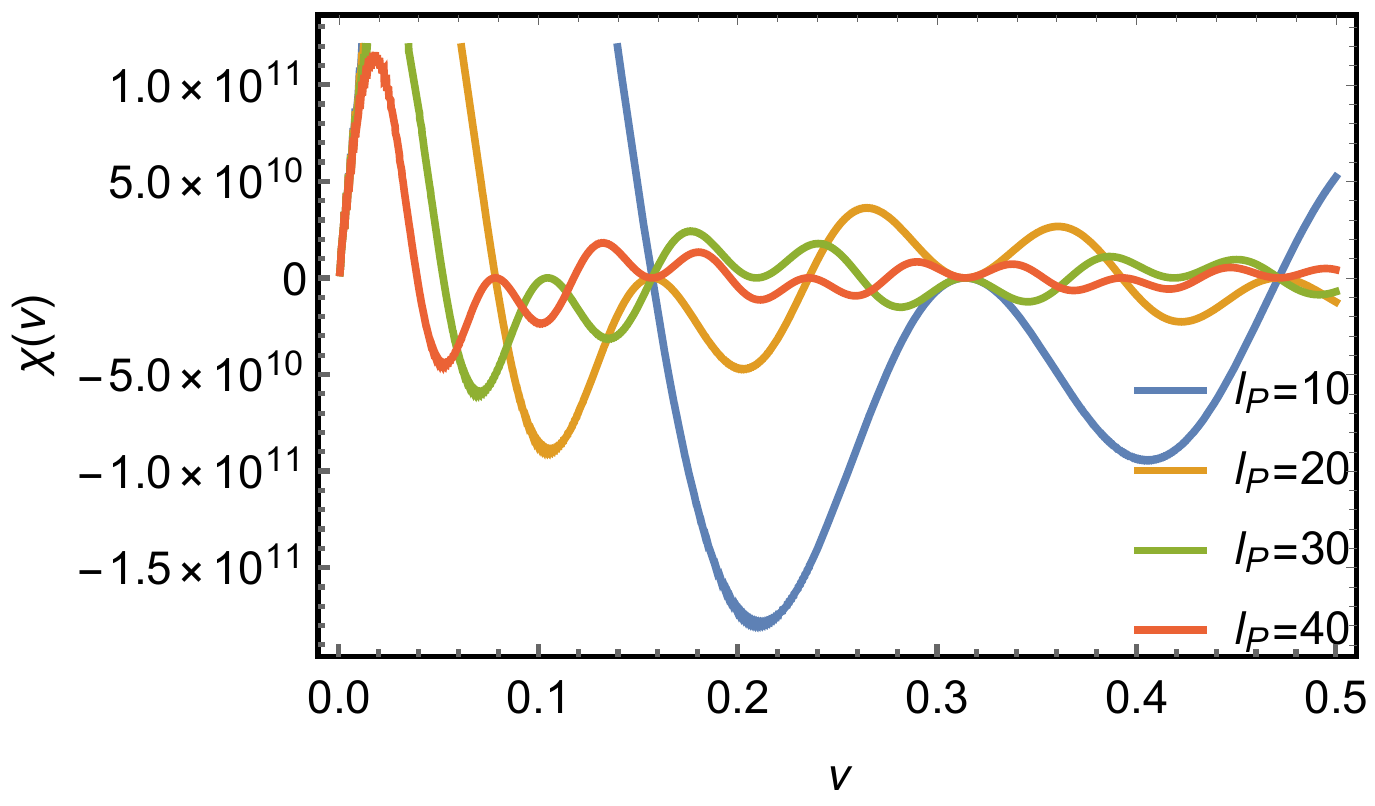}
\caption{The spectral radiance to inflationary (top left), electroweak (top right) and cosmic microwave back ground (the bottom) periods of the universe for different values of $l_{P}$.}
\label{spectralradiance-l}
\end{figure}

\begin{figure}[tbh]
\centering
\includegraphics[width=8cm,height=5cm]{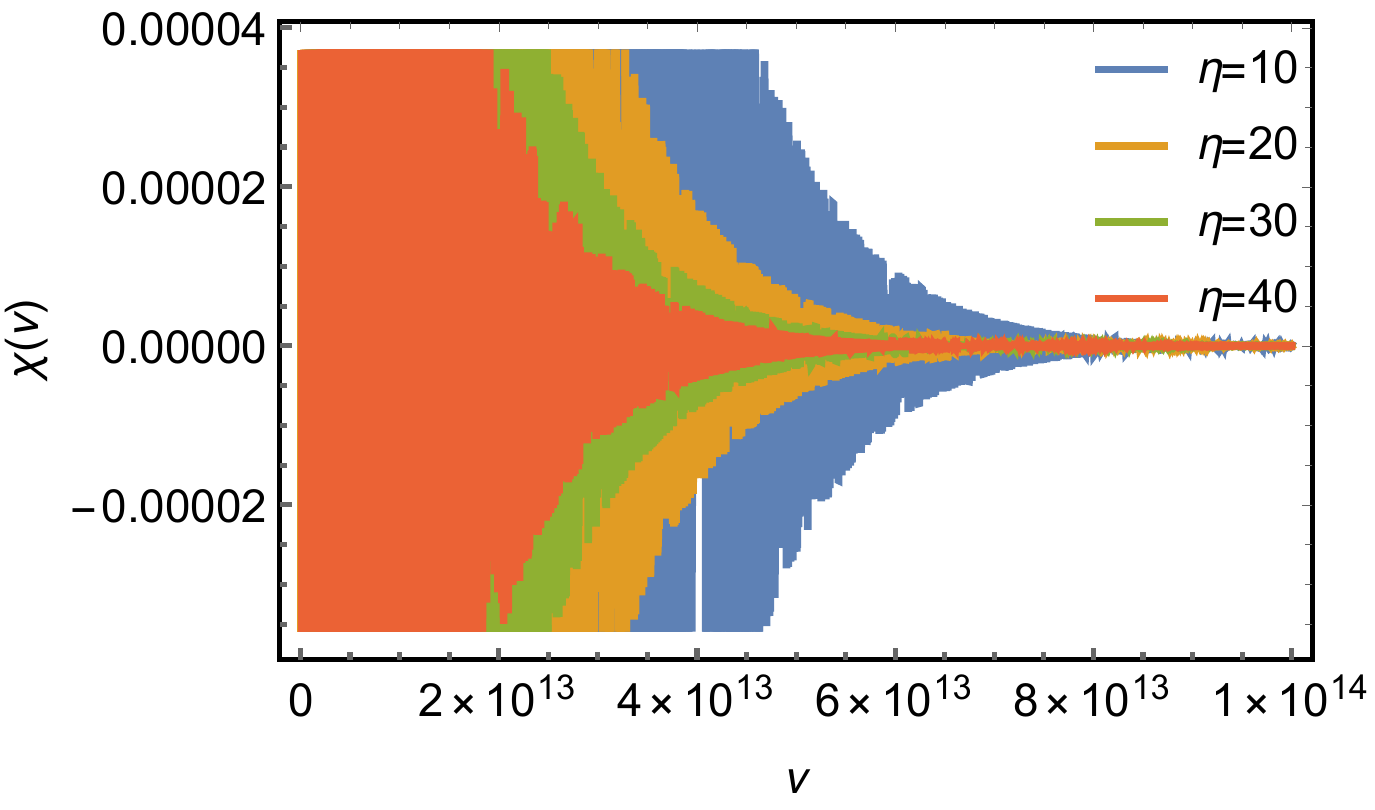}
\includegraphics[width=8cm,height=5cm]{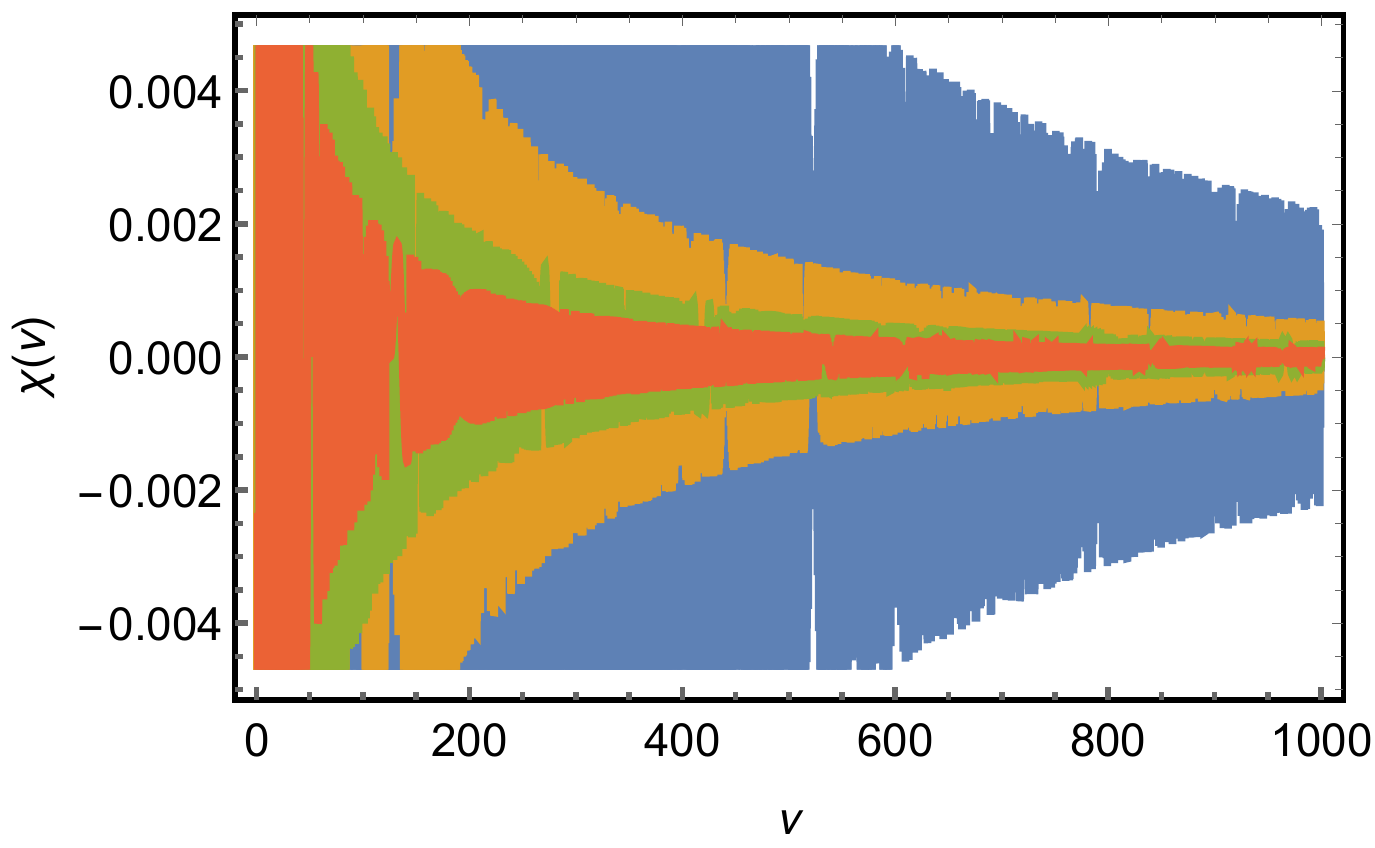}
\includegraphics[width=8cm,height=5cm]{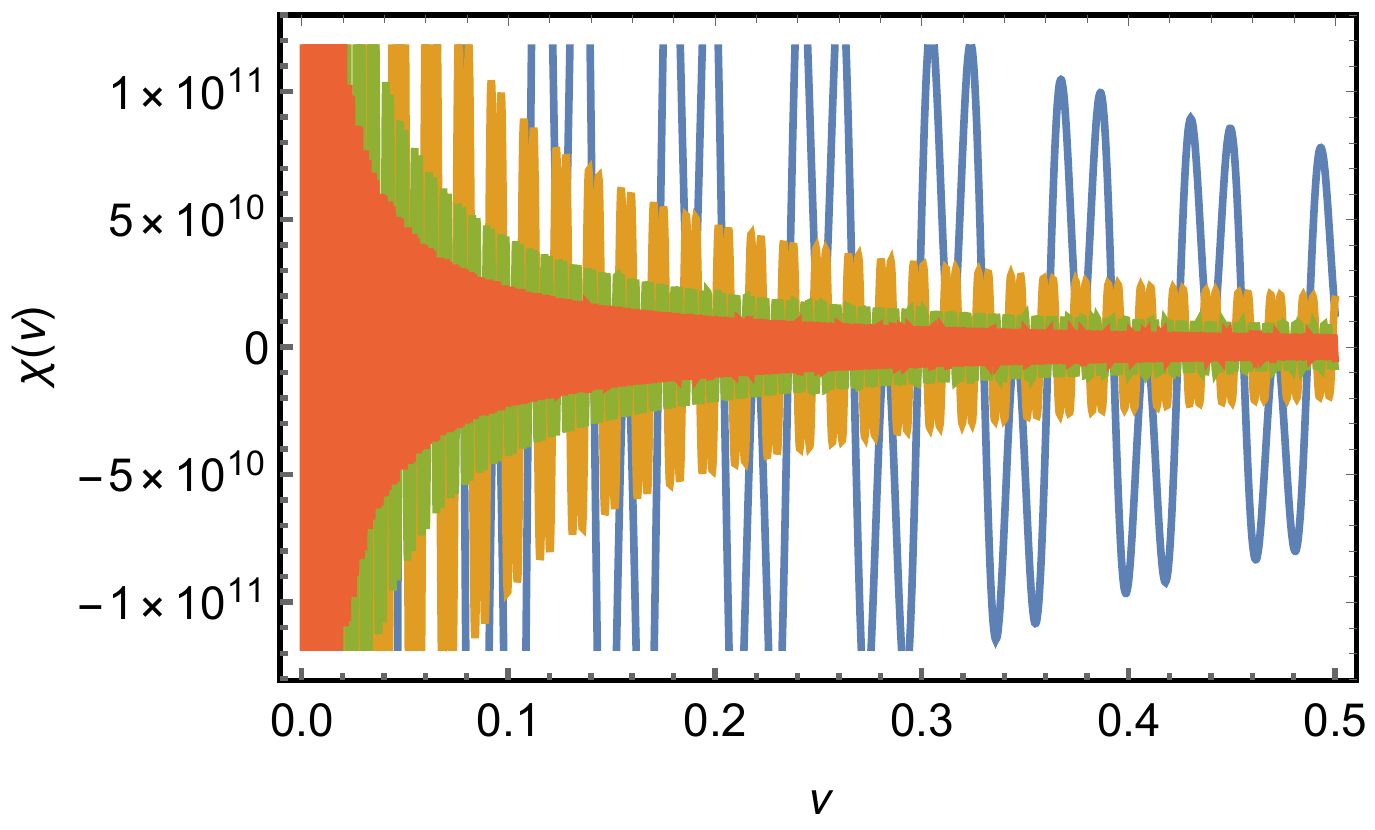}
\caption{The spectral radiance to inflationary (top left), electroweak (top right) and cosmic microwave back ground (the bottom) periods of the universe for different values of $\eta$.}
\label{spectralradiance-eta}
\end{figure}

We see that the plots displayed in Figs. \ref{acessiblestates-inflation}, and \ref{acessiblestates-electroweak} have a similar behavior differing from each other mainly in the energy scale. On the other hand, in the CMB scenario exhibited in Fig. \ref{acessiblestates-cmb}, the accessible states of the system turn out to be sheets having, therefore, their behavior less complex than the two previous cases. Here, for a better comprehension, let us introduce a generic definition of the partition function for an indistinguishable spinless gas \cite{greiner2012thermodynamics}: 
\ie
Z(T,\Gamma,N) = \frac{1}{N!h^{3N}} \int \mathrm{d}q^{3N}\mathrm{d}p^{3N} e^{-\beta H(q,p)}  \equiv \int \mathrm{d}E \,\Omega(E) e^{-\beta E}, \label{partti1}
\fe
where $q$ is the generalized coordinates, $p$ is the generalized momenta, $N$ is the number of particles, $H$ is the Hamiltonian of the system, $h$ is the Planck's constant. Nevertheless, Eq. (\ref{partti1}), does not account for the spin of the particles that we are dealing with. In our case, we have considered photons (bosons) which reads: 
\ie
\mathrm{ln}[Z] = \int \mathrm{d}E \,\Omega(E) \mathrm{ln} [ 1- e^{-\beta E}],
\fe
where the factor $\mathrm{ln} [ 1- e^{-\beta E}]$ accounts for the Bose-Einstein statistics.
Next, after using Eq. (\ref{accessiblestates}), the partition function can be written in a straightforward manner as
\ie
\begin{split}
\mathrm{ln}[Z(\beta,\eta,l_{P})]  = &-\frac{1}{\pi^{2}} \int^{\infty}_{0} \frac{1}{\eta^{2} l^{2}_{P}} \sin^{2}(\eta^{2} l_{P} E)\cos(\eta^{2} l_{P} E) \,\mathrm{ln}\left( 1 - e^{-\beta E}  \right) \,\mathrm{d}E \\
= & -\frac{\pi^{2}}{\eta^{2}l_{P}^{2}}\left\{\frac{\beta e^{\beta E +i\eta^2 l_{P} E } \, _2F_1\left(1,\frac{i l_{P} \eta^2+\beta}{\beta};\frac{i l_{P} \eta^2+2 \beta}{\beta};e^{\beta E }\right)}{-8 \eta^4 l_{P}^2+8 i \beta  \eta^2 l_{P}} \right.\\
& \left. + \frac{1}{72 l^{2}_{P} \eta^{4}}\left[+\frac{3 i \beta  \eta^2 l_{P} e^{E  \left(\beta +3 i \eta^2 l_{P}\right)} \, _2F_1\left(1,\frac{3 i l_{P} \eta^2}{\beta }+1;\frac{3 i l_{P} \eta^2}{\beta }+2;e^{\beta  E }\right)}{\beta +3 i \eta^2 l_{P}} \right. \right. \\
&\left. \left. -\frac{9 \beta  \eta^2 l_{P} e^{E  \left(\beta -i \eta^2 l_{P}\right)} \, _2F_1\left(1,1-\frac{i l_{P} \eta^2}{\beta };2-\frac{i l_{P} \eta^2}{\beta };e^{\beta  E }\right)}{\eta^2 l_{P}+i \beta } \right. \right.\\
&\left. \left. -\frac{3 i \beta  \eta^2 l_{P} e^{E  \left(\beta -3 i \eta^2 l_{P}\right)} \, _2F_1\left(1,1-\frac{3 i l_{P} \eta^2}{\beta };2-\frac{3 i l_{P} \eta^2}{\beta };e^{\beta  E }\right)}{\beta -3 i \eta^2 l_{P}} \right.\right. \\
& \left.\left. -18 \beta  \cos \left(\eta^2 l_{P} E \right)+2 \beta  \cos \left(3 \eta^2 l_{P} E \right)+18 \eta^2 l_{P} \mathrm{ln} \left(1-e^{-\beta  E }\right) \sin \left(\eta^2 l_{P} E \right) \right.\right. \\
&\left.\left. -6 \eta^2 l_{P} \mathrm{ln} \left(1-e^{-\beta  E }\right) \sin \left(3 \eta^2 l_{P} E \right) \right] \right\} \Bigg\rvert^{\infty}_{0} \\
& = - \frac{\pi ^2 \left\{8 \beta +3 \pi  \eta ^2 l_{P} \left[\coth \left(\frac{3 \pi  \eta ^2 l_{P}}{\beta }\right)-3 \coth \left(\frac{\pi  \eta ^2 l_{P}}{\beta }\right)\right]\right\}}{72  \eta ^6 l_{P}^4},
\end{split}
\fe
where $\beta = 1/ \kappa_{B} T$ and $F_{1}$ is the hypergeometric function. Here, with the partition function, all quantities of interest can be derived using the definitions below:
\ie
\begin{split}
 & F(\beta,\eta,l_{P})=-\frac{1}{\beta} \mathrm{ln}\left[Z(\beta,\eta,l_{P})\right], \\
 & U(\beta,\eta,l_{P})=-\frac{\partial}{\partial\beta} \mathrm{ln}\left[Z(\beta,\eta,l_{P})\right], \\
 & S(\beta,\eta,l_{P})=k_B\beta^2\frac{\partial}{\partial\beta}F(\beta,\eta,l_{P}), \\
 & C_V(\beta,\eta,l_{P})=-k_B\beta^2\frac{\partial}{\partial\beta}U(\beta,\eta,l_{P}).
\label{properties}
\end{split}
\fe  
It is worth mentioning that regarding the thermal aspects, other previous studies were derived in different scenarios \cite{oliveira2019thermodynamic,oliveira2020thermodynamic,ikot2021klein,ikot2021approximate,reis2020does,reis2021fermions} as well. In other to accomplish our investigation, we begin to analyze the mean energy which takes the form
\ie
\begin{split}
& U(\beta,\eta,l_{P})  = \frac{1}{\pi^{2}} \int^{\infty}_{0} \frac{1}{\eta^{2} l^{2}_{P}} \sin^{2}(\eta^{2} l_{P} E)\cos(\eta^{2} l_{P} E) \frac{e^{-\beta E}}{1 - e^{-\beta E}} \,\mathrm{d}E \\
= &\frac{1}{72 l^{4}_{P}\pi^{2}\eta^{6}} \\
& \times \left\{ \left(-9 \sin \left(E \eta ^2 l_{P}\right)-9 i \cos \left(E \eta ^2 l_{P}\right)\right) \left[E \eta ^2 l_{P} \, _2F_1\left(1,-\frac{i l_{P} \eta ^2}{\beta };1-\frac{i l_{P} \eta ^2}{\beta };\cosh (\beta  E) +\sinh (\beta  E)\right) \right.\right.\\
& \left.\left.-i \, _3F_2\left(1,-\frac{i l_{P} \eta ^2}{\beta },-\frac{i l_{P} \eta ^2}{\beta };1-\frac{i l_{P} \eta ^2}{\beta },1-\frac{i l_{P} \eta ^2}{\beta };\cosh (\beta  E)+\sinh (\beta  E)\right)\right]+\left(-9 \sin \left(E \eta ^2 l_{P}\right) \right.\right.\\
& \left.\left. +9 i \cos \left(E \eta ^2 l_{P}\right)\right) \left[i \, _3F_2\left(1,\frac{i l_{P} \eta ^2}{\beta },\frac{i l_{P} \eta ^2}{\beta };\frac{i l_{P} \eta ^2}{\beta }+1,\frac{i l_{P} \eta ^2}{\beta }+1;\cosh (\beta  E)+\sinh (\beta  E)\right) \right.\right.\\
& \left.\left. +E \eta ^2 l_{P} \, _2F_1\left(1,\frac{i l_{P} \eta ^2}{\beta };\frac{i l_{P} \eta ^2}{\beta }+1;\cosh (\beta  E)+\sinh (\beta  E)\right)\right] \right. \\
& \left. +\left(\cos \left(3 E \eta ^2 l_{P}\right)+i \sin \left(3 E \eta ^2 l_{P}\right)\right) \right.\\ 
& \left. \times \left[\, _3F_2\left(1,\frac{3 i l_{P} \eta ^2}{\beta },\frac{3 i l_{P} \eta ^2}{\beta };\frac{3 i l_{P} \eta ^2}{\beta }+1,\frac{3 i l_{P} \eta ^2}{\beta }+1;\cosh (\beta  E)+\sinh (\beta  E)\right) \right.\right.\\
& \left.\left. -3 i E \eta ^2 l_{P} \, _2F_1\left(1,\frac{3 i l_{P} \eta ^2}{\beta };\frac{3 i l_{P} \eta ^2}{\beta }+1;\cosh (\beta  E)+\sinh (\beta  E)\right)\right]+ \right. \\
& \left. \left(\cos \left(3 E \eta ^2 l_{P}\right)-i \sin \left(3 E \eta ^2 l_{P}\right)\right) \right.\\
& \left. \times \left[\, _3F_2\left(1,-\frac{3 i l_{P} \eta ^2}{\beta },-\frac{3 i l_{P} \eta ^2}{\beta };1-\frac{3 i l_{P} \eta ^2}{\beta },1-\frac{3 i l_{P} \eta ^2}{\beta };\cosh (\beta  E)+\sinh (\beta  E)\right) \right.\right. \\
& \left.\left. +3 i E \eta ^2 l_{P} \, _2F_1\left(1,-\frac{3 i l_{P} \eta ^2}{\beta };1-\frac{3 i l_{P} \eta ^2}{\beta };\cosh (\beta  E)+\sinh (\beta  E)\right)\right]  \right\} \Bigg\rvert^{\infty}_{0} = 0. \label{mean-energy}
\end{split}
\fe
Here, notice that this expression displays a fascinating effect: after considering the limits of integration, the mean energy does not yield any contribution besides the trivial one. Moreover, from Eq. (\ref{mean-energy}), we immediately obtain the spectral radiance given by 
\ie
\chi(\eta,l_{P},\nu)  =  \frac{1}{\pi^{2}\eta^{2} l^{2}_{P}} \sin^{2}(\eta^{2} l_{P} h \nu)\cos(\eta^{2} l_{P} h \nu) \frac{e^{-\beta h \nu}}{1 - e^{-\beta h \nu}},
\fe
where we have considered $E=h\nu$, being $h$ the Planck constant\footnote{For the sake of simplicity, we shall consider $\kappa_{B}=h=1$.} and $\nu$ the frequency. The plots of such a radiance are displayed in Figs. \ref{spectralradiance-l}, and \ref{spectralradiance-eta}. Specifically,  for the Fig. \ref{spectralradiance-l}, we examine how the parameter $l_{P}$ modifies the accessible states of the system. Further, in Fig. \ref{spectralradiance-eta}, we perform the same examination for the parameter $\eta$ though. Intriguing, for both scenarios, the behavior of the spectral radiance turned out to be unusual. This occurs because they exhibited a fluctuation between positive and negative values of frequency as a mirror. Moreover, this might indicate instability from the starting model mainly perhaps due to the periodic characteristic of the dispersion relation of Eq. (\ref{sin}).

\begin{figure}[tbh]
\centering
\includegraphics[width=6cm,height=4cm]{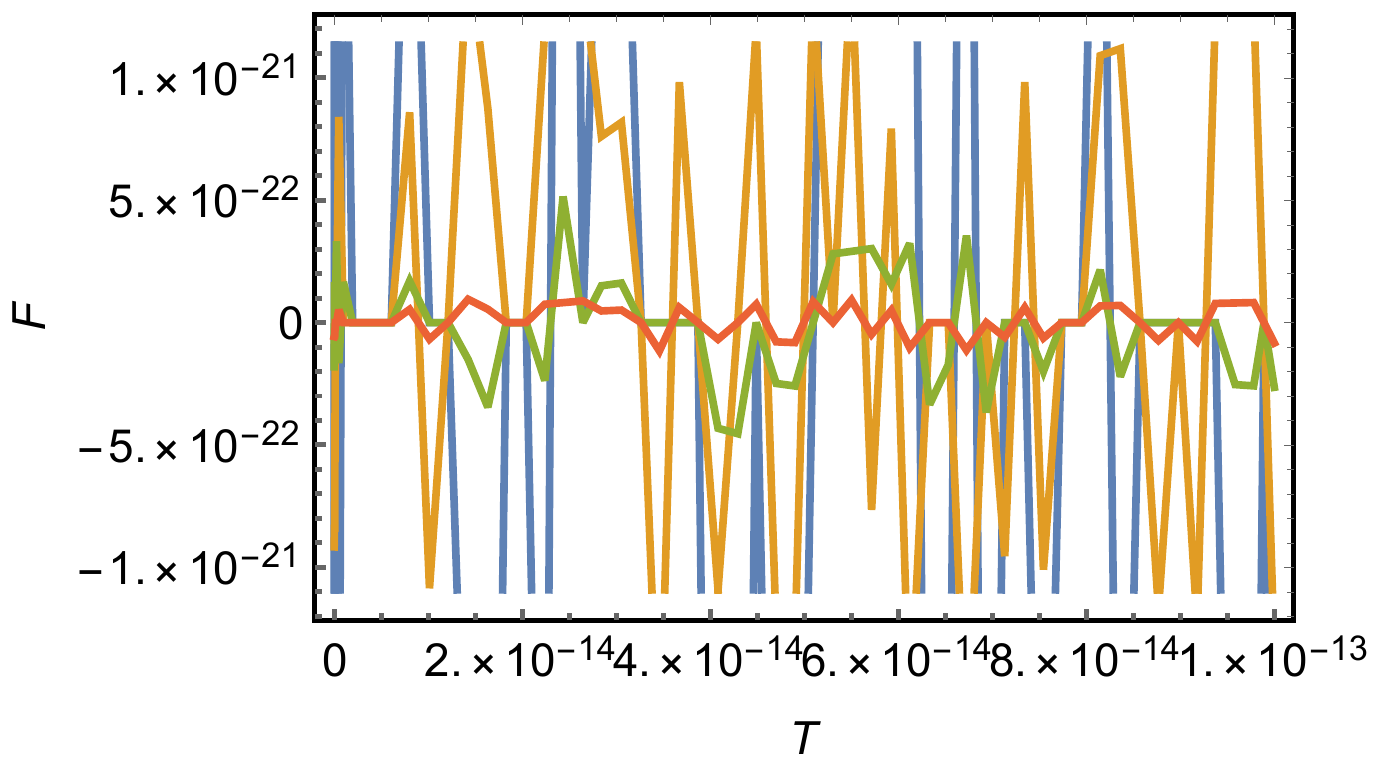}
\includegraphics[width=6cm,height=4cm]{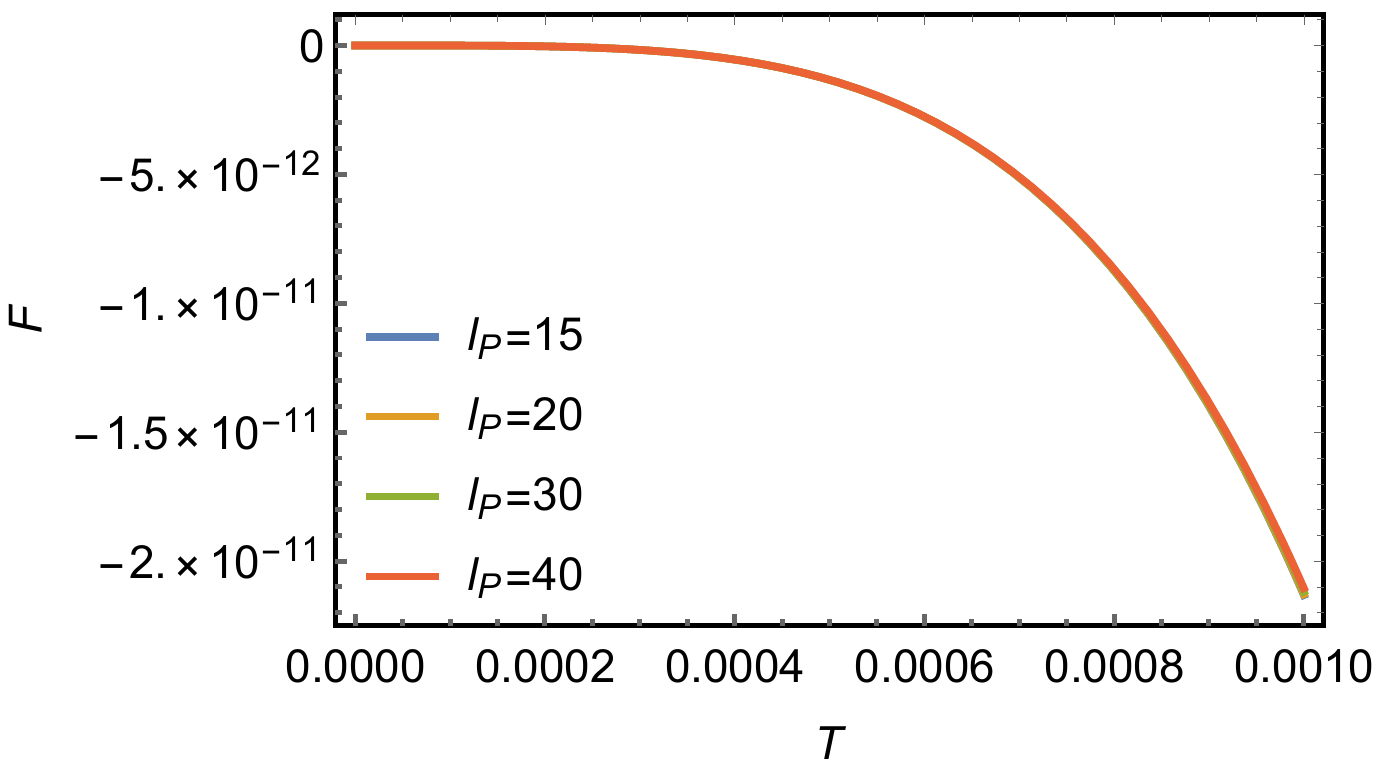}
\includegraphics[width=6cm,height=4cm]{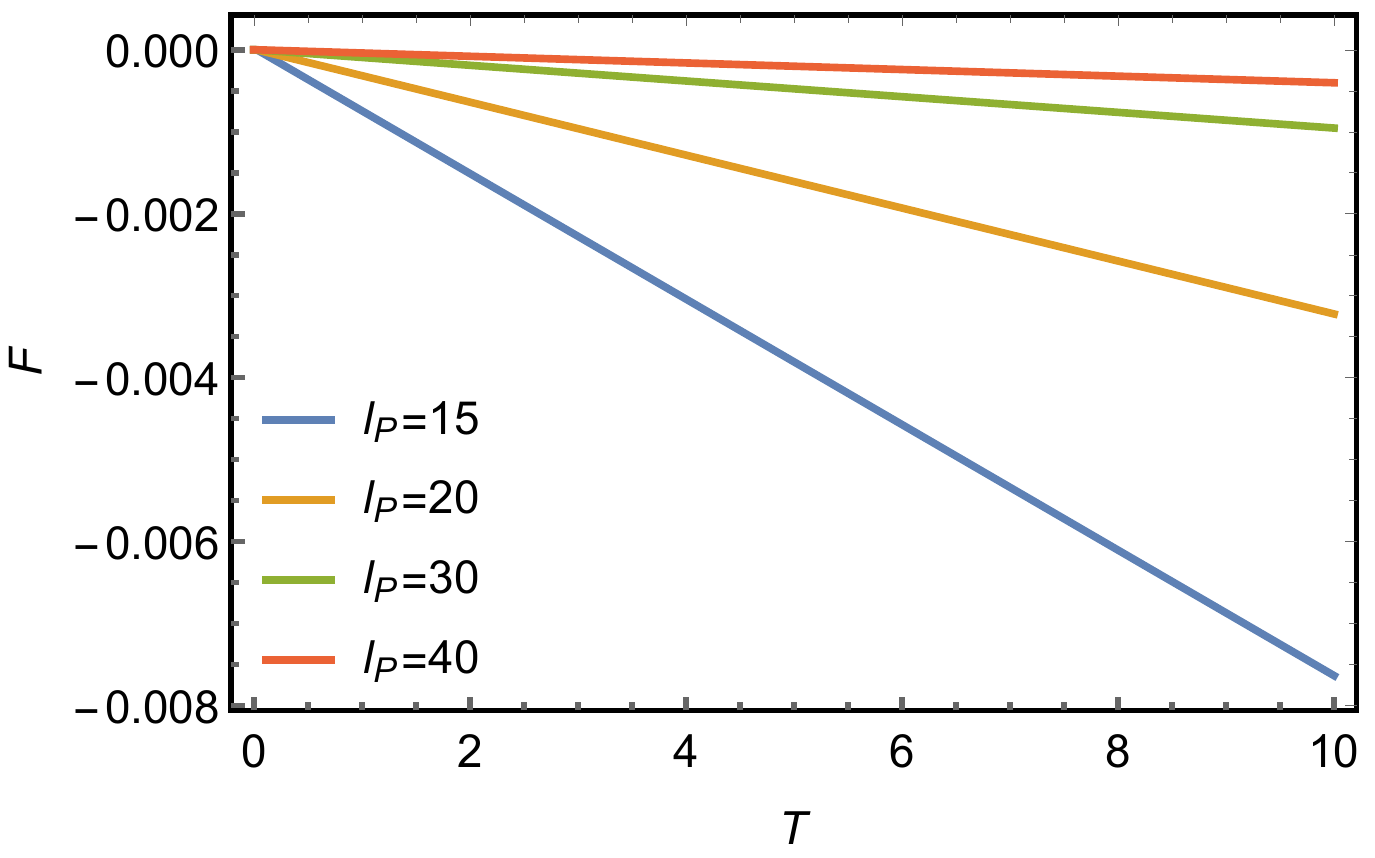}
\includegraphics[width=6cm,height=4cm]{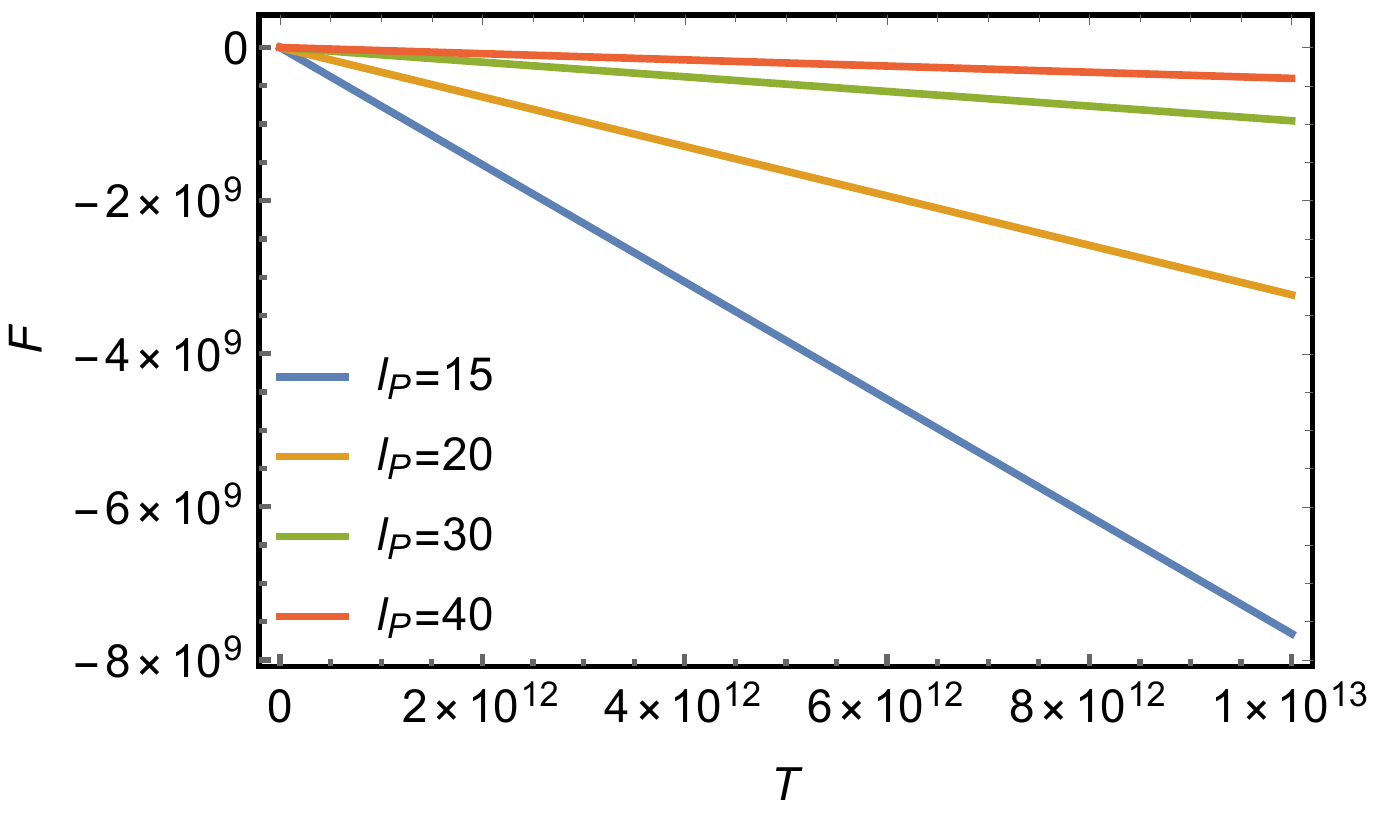}
\caption{The Helmholtz free energy for different values of $l_{P}$.}
\label{Helmholtz}
\end{figure}

\begin{figure}[tbh]
\centering
\includegraphics[width=6cm,height=4cm]{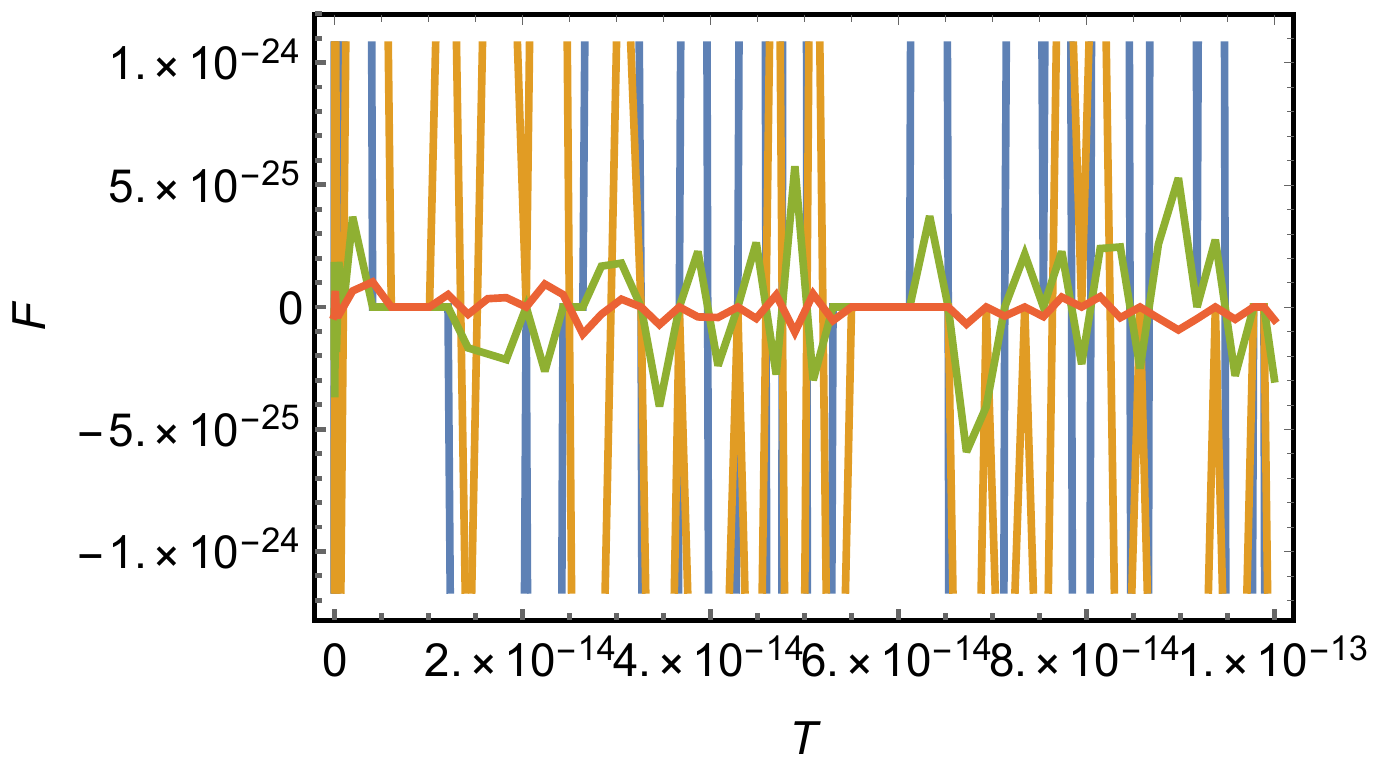}
\includegraphics[width=6cm,height=4cm]{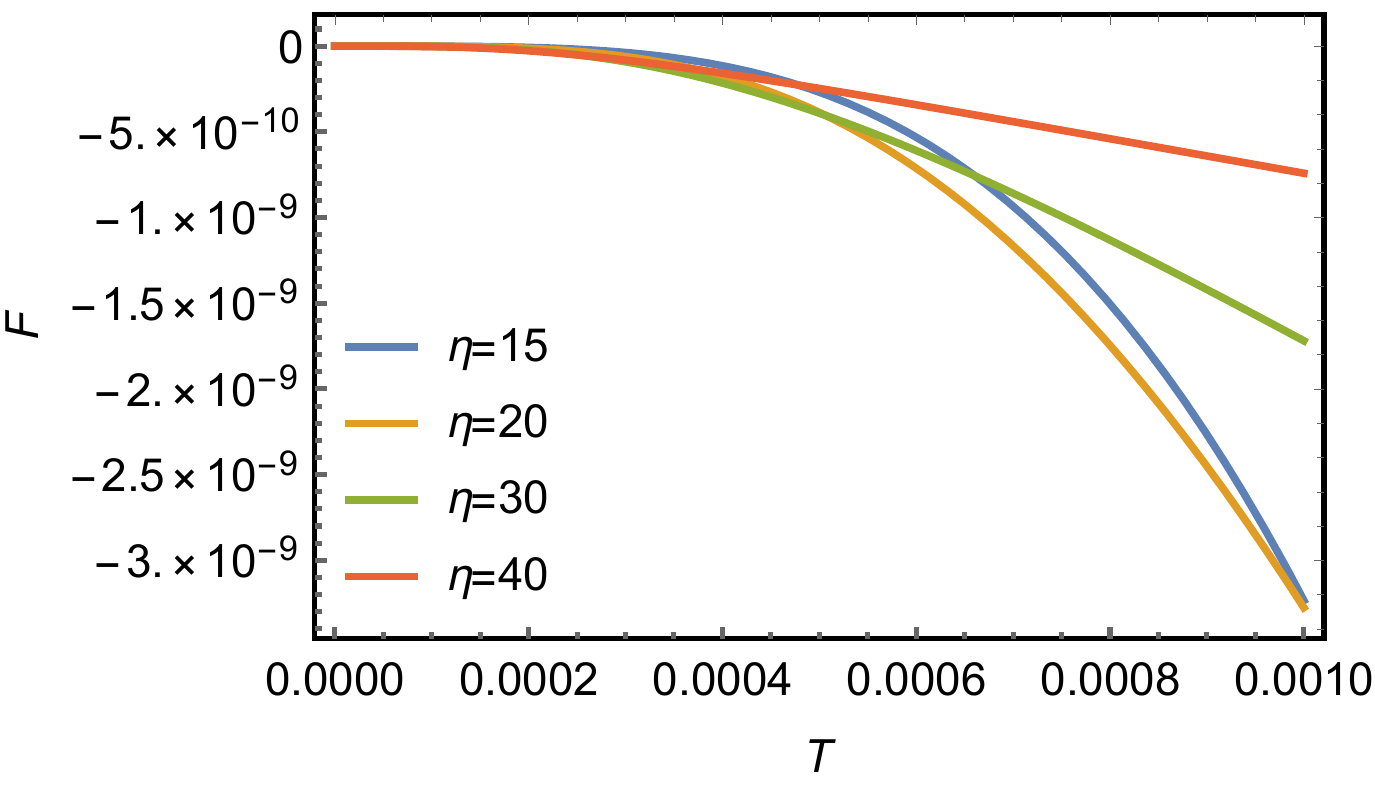}
\includegraphics[width=6cm,height=4cm]{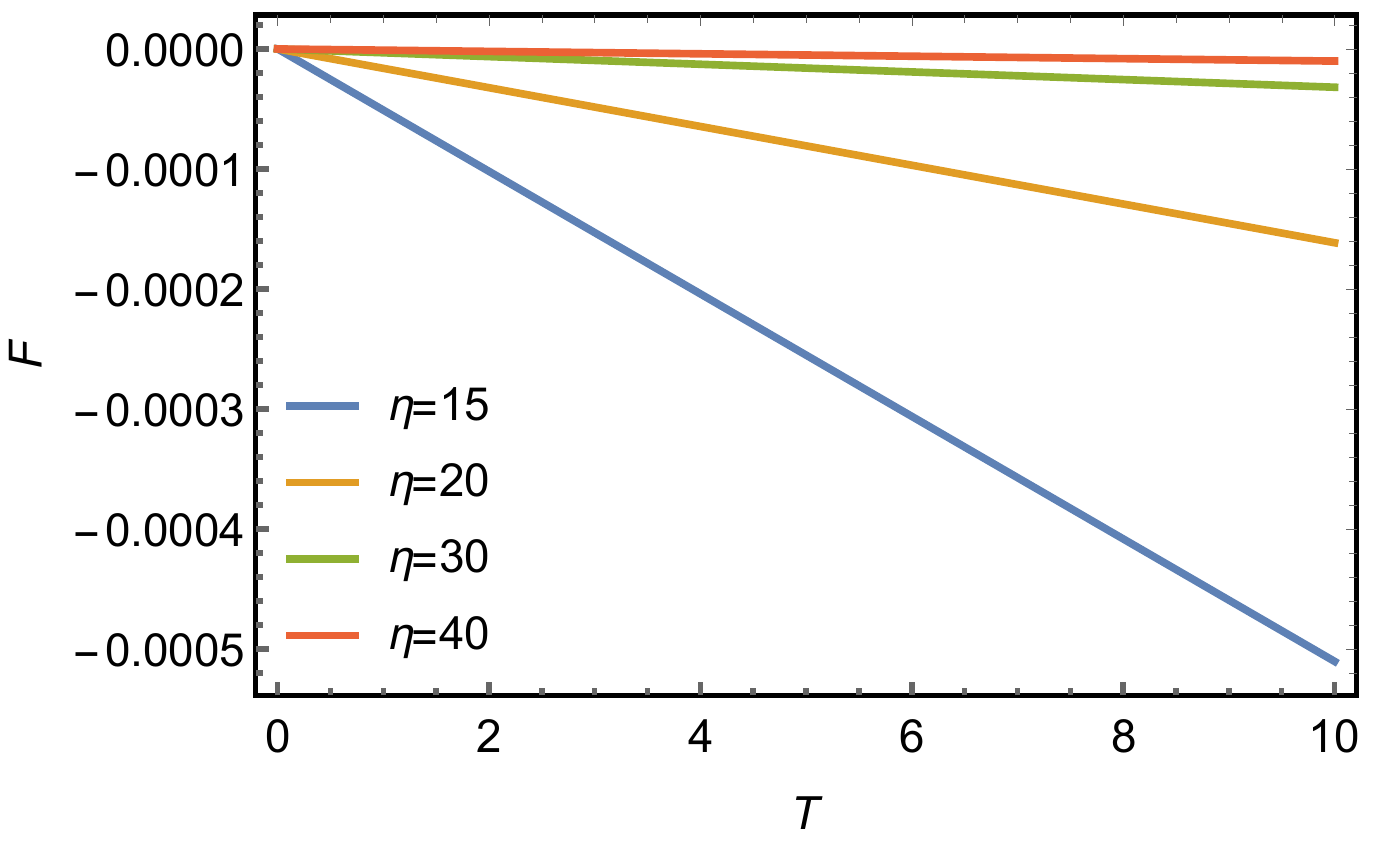}
\includegraphics[width=6cm,height=4cm]{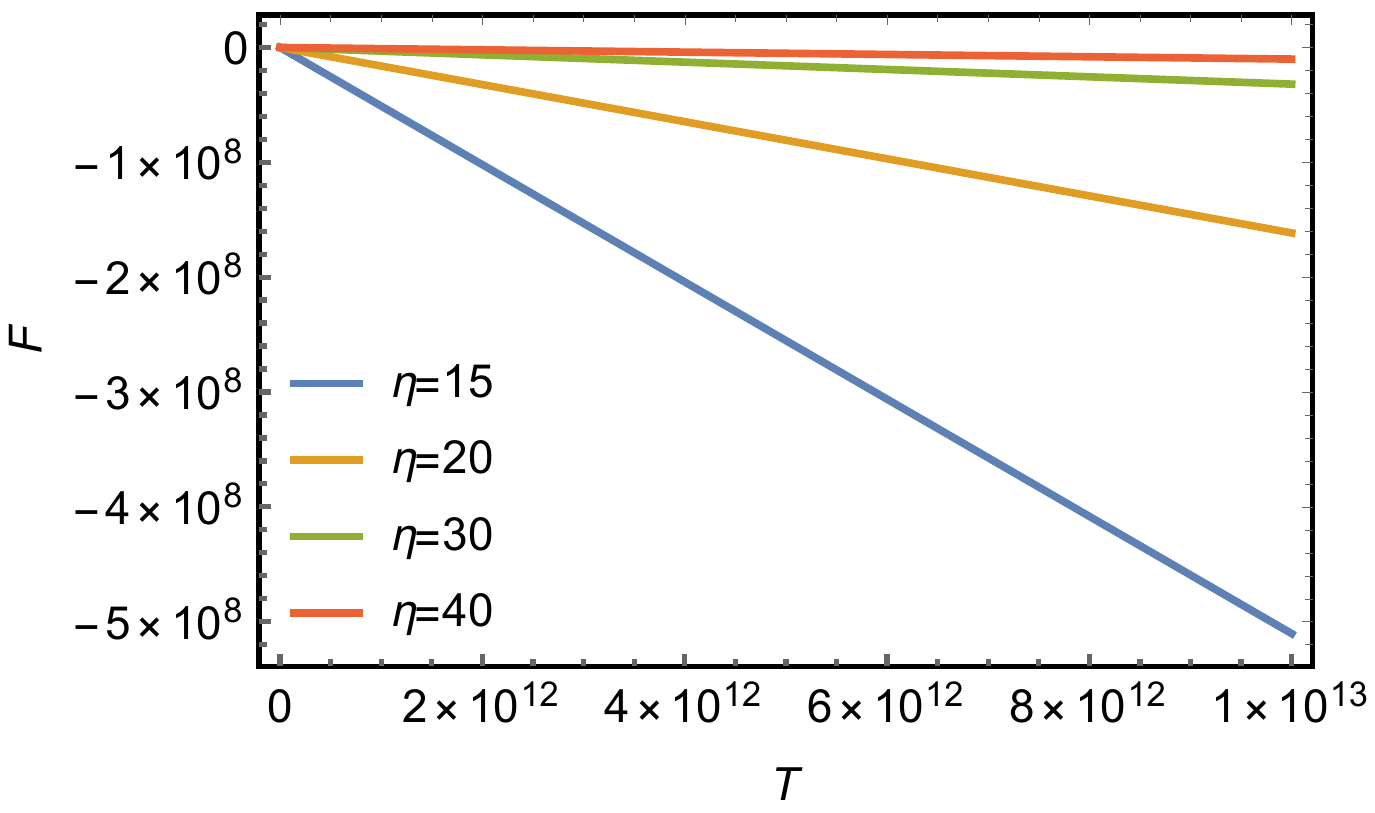}
\caption{The Helmholtz free energy for different values of $\eta$.}
\label{Helmholtz2}
\end{figure}

Next, we calculate the Helmholtz free energy
\ie
\begin{split}
& F(\beta,\eta,l_{P})  =  \frac{1}{\beta\pi^{2}} \int_{0}^{\infty} \frac{1}{\eta^{2} l^{2}_{P}} \sin^{2}(\eta^{2} l_{P} E)\cos(\eta^{2} l_{P} E) \,\mathrm{ln}\left( 1 - e^{-\beta E}  \right) \,\mathrm{d}E\\
= & \frac{\pi^{2}}{\beta \eta^{2}l_{P}^{2}}\left\{\frac{\beta e^{\beta E +i\eta^2 l_{P} E } \, _2F_1\left(1,\frac{i l_{P} \eta^2+\beta}{\beta};\frac{i l_{P} \eta^2+2 \beta}{\beta};e^{\beta E }\right)}{-8 \eta^4 l_{P}^2+8 i \beta  \eta^2 l_{P}} \right.\\
& \left. + \frac{1}{72 l^{2}_{P} \eta^{4}}\left[+\frac{3 i \beta  \eta^2 l_{P} e^{E  \left(\beta +3 i \eta^2 l_{P}\right)} \, _2F_1\left(1,\frac{3 i l_{P} \eta^2}{\beta }+1;\frac{3 i l_{P} \eta^2}{\beta }+2;e^{\beta  E }\right)}{\beta +3 i \eta^2 l_{P}} \right. \right. \\
&\left. \left. -\frac{9 \beta  \eta^2 l_{P} e^{E  \left(\beta -i \eta^2 l_{P}\right)} \, _2F_1\left(1,1-\frac{i l_{P} \eta^2}{\beta };2-\frac{i l_{P} \eta^2}{\beta };e^{\beta  E }\right)}{\eta^2 l_{P}+i \beta } \right. \right.\\
&\left. \left. -\frac{3 i \beta  \eta^2 l_{P} e^{E  \left(\beta -3 i \eta^2 l_{P}\right)} \, _2F_1\left(1,1-\frac{3 i l_{P} \eta^2}{\beta };2-\frac{3 i l_{P} \eta^2}{\beta };e^{\beta  E }\right)}{\beta -3 i \eta^2 l_{P}} \right.\right. \\
& \left.\left. -18 \beta  \cos \left(\eta^2 l_{P} E \right)+2 \beta  \cos \left(3 \eta^2 l_{P} E \right)+18 \eta^2 l_{P} \mathrm{ln} \left(1-e^{-\beta  E }\right) \sin \left(\eta^2 l_{P} E \right) \right.\right. \\
&\left.\left. -6 \eta^2 l_{P} \mathrm{ln} \left(1-e^{-\beta  E }\right) \sin \left(3 \eta^2 l_{P} E \right) \right] \right\} \Bigg\rvert^{\infty}_{0} \\
& = \frac{\pi ^2 \left\{8 \beta +3 \pi  \eta ^2 l_{P} \left[\coth \left(\frac{3 \pi  \eta ^2 l_{P}}{\beta }\right)-3 \coth \left(\frac{\pi  \eta ^2 l_{P}}{\beta }\right)\right]\right\}}{72 \beta  \eta ^6 l_{P}^4}.
\end{split}
\fe
Differently from the result encountered in the mean energy, for the case of the Helmholtz free energy, we arrive at a nontrivial solution after substituting the limits of integration. The behavior of such thermodynamic function is displayed in Fig. \ref{Helmholtz} and Fig. \ref{Helmholtz2}. In Fig. \ref{Helmholtz}, we plot such thermal function for different values of $l_{P}$. Likewise, in Fig. \ref{Helmholtz2}, we study the behavior for rather distinct values of the parameter $\eta$.

In general, these curves totally agree with the known results, except in the case when we consider the temperature of the cosmic microwave background, i.e., T = $10^{-13}$ GeV. The graphics indicate that there exist oscillations between positive and negative values. In this case, it might indicate that our system is close to the bouncing region and, therefore, some disturbing/unusual consequences might take place.    

Furthermore, one important thermodynamic function to be taken into account is the entropy. In this sense,
\ie
\begin{split}
S(\beta,\eta,l_{P})  & =  \frac{\kappa_{B}}{\pi^{2}} \int^{\infty}_{0} \frac{1}{\eta^{2} l^{2}_{P}} \sin^{2}(\eta^{2} l_{P} E)\cos(\eta^{2} l_{P} E) \,\mathrm{ln}\left( 1 - e^{-\beta E}  \right) \,\mathrm{d}E \\
& + \frac{\kappa_{B} \beta}{\pi^{2}} \int^{\infty}_{0} \frac{E}{\eta^{2} l^{2}_{P}} \sin^{2}(\eta^{2} l_{P} E)\cos(\eta^{2} l_{P} E) \frac{e^{-\beta E}}{1-e^{-\beta E}} 
\,\mathrm{d}E\\
&=  \frac{1}{72 \pi ^2 \eta ^6 l_{P}^4}  \left\{ 9 \beta  e^{i E \eta ^2 l_{P}} \left[\left(1+i E \eta ^2 l_{P}\right) \, _2F_1\left(1,\frac{i l_{P} \eta ^2}{\beta };\frac{i l_{P} \eta ^2}{\beta }+1;e^{\beta  E}\right)  \right. \right. \\
& \left. \left. -_3F_2\left(1,\frac{i l_{P} \eta ^2}{\beta },\frac{i l_{P} \eta ^2}{\beta };\frac{i l_{P} \eta ^2}{\beta }+1,\frac{i l_{P} \eta ^2}{\beta }+1;e^{\beta  E}\right)\right] \right.\\
&\left. +\beta  e^{3 i E \eta ^2 l_{P}} \left[\, _3F_2\left(1,\frac{3 i l_{P} \eta ^2}{\beta },\frac{3 i l_{P} \eta ^2}{\beta };\frac{3 i l_{P} \eta ^2}{\beta }+1,\frac{3 i l_{P} \eta ^2}{\beta }+1;e^{\beta  E}\right) \right. \right.\\
& \left. \left.+ \left(-1-3 i E \eta ^2 l_{P}\right) \, _2F_1\left(1,\frac{3 i l_{P} \eta ^2}{\beta };\frac{3 i l_{P} \eta ^2}{\beta } + 1;e^{\beta  E}\right)\right] \right. \\
& \left. +9 \beta  e^{-i E \eta ^2 l_{P}} \left[\left(1-i E \eta ^2 l_{P}\right) \, _2F_1\left(1,-\frac{i l_{P} \eta ^2}{\beta };1-\frac{i l_{P} \eta ^2}{\beta };e^{\beta  E}\right) \right.\right.\\
&\left.\left.- \, _3F_2\left(1,-\frac{i l_{P} \eta ^2}{\beta },-\frac{i l_{P} \eta ^2}{\beta };1-\frac{i l_{P} \eta ^2}{\beta },1-\frac{i l_{P} \eta ^2}{\beta };e^{\beta  E}\right)\right] \right.\\
& \left.+ \beta  e^{-3 i E \eta ^2 l_{P}} \left[\, _3F_2\left(1,-\frac{3 i l_{P} \eta ^2}{\beta },-\frac{3 i l_{P} \eta ^2}{\beta };1-\frac{3 i l_{P} \eta ^2}{\beta },1-\frac{3 i l_{P} \eta ^2}{\beta };e^{\beta  E}\right) \right. \right.\\
& \left.\left.+\left(-1+3 i E \eta ^2 l_{P}\right) \, _2F_1\left(1,-\frac{3 i l_{P} \eta ^2}{\beta };1-\frac{3 i l_{P} \eta ^2}{\beta };e^{\beta  E}\right)\right]\right.\\
& \left.-24 \eta ^2 l_{P} \mathrm{ln} \left(e^{\beta  E}-1\right) \sin ^3\left(E \eta ^2 l_{P}\right)-18 \eta ^2 l_{P} \left[\mathrm{ln} \left(1-e^{-\beta  E}\right) \right.\right.\\
& \left. \left. -\mathrm{ln} \left(e^{\beta  E}-1\right)\right] \sin \left(E \eta ^2 l_{P}\right)+6 \eta ^2 l_{P} \left(\mathrm{ln} \left(1-e^{-\beta  E}\right)  \right.\right. \\
& \left.\left. - \mathrm{ln} \left(e^{\beta  E}-1\right)\right) \sin \left(3 E \eta ^2 l_{P}\right) \right\} \Bigg\rvert^{\infty}_{0} = \frac{3 \pi  \eta ^2 l_{P} \left[3 \coth \left(\frac{\pi  \eta ^2  l_{P}}{\beta }\right)-\coth \left(\frac{3 \pi  \eta ^2 l_{P}}{\beta }\right)\right]-8 \beta }{72 \pi ^2 \eta ^6 l_{P}^4}.
\end{split}
\fe

\begin{figure}[tbh]
\centering
\includegraphics[width=6cm,height=4cm]{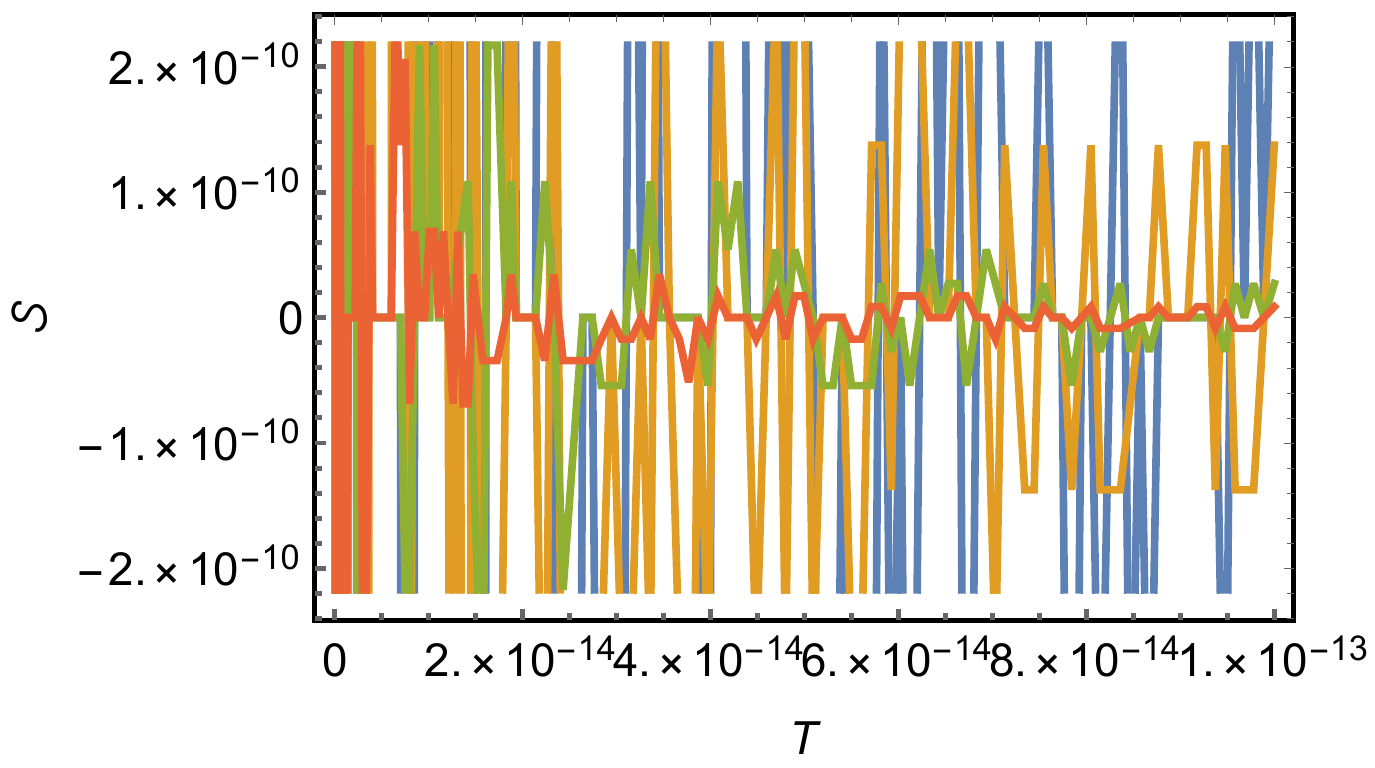}
\includegraphics[width=6cm,height=4cm]{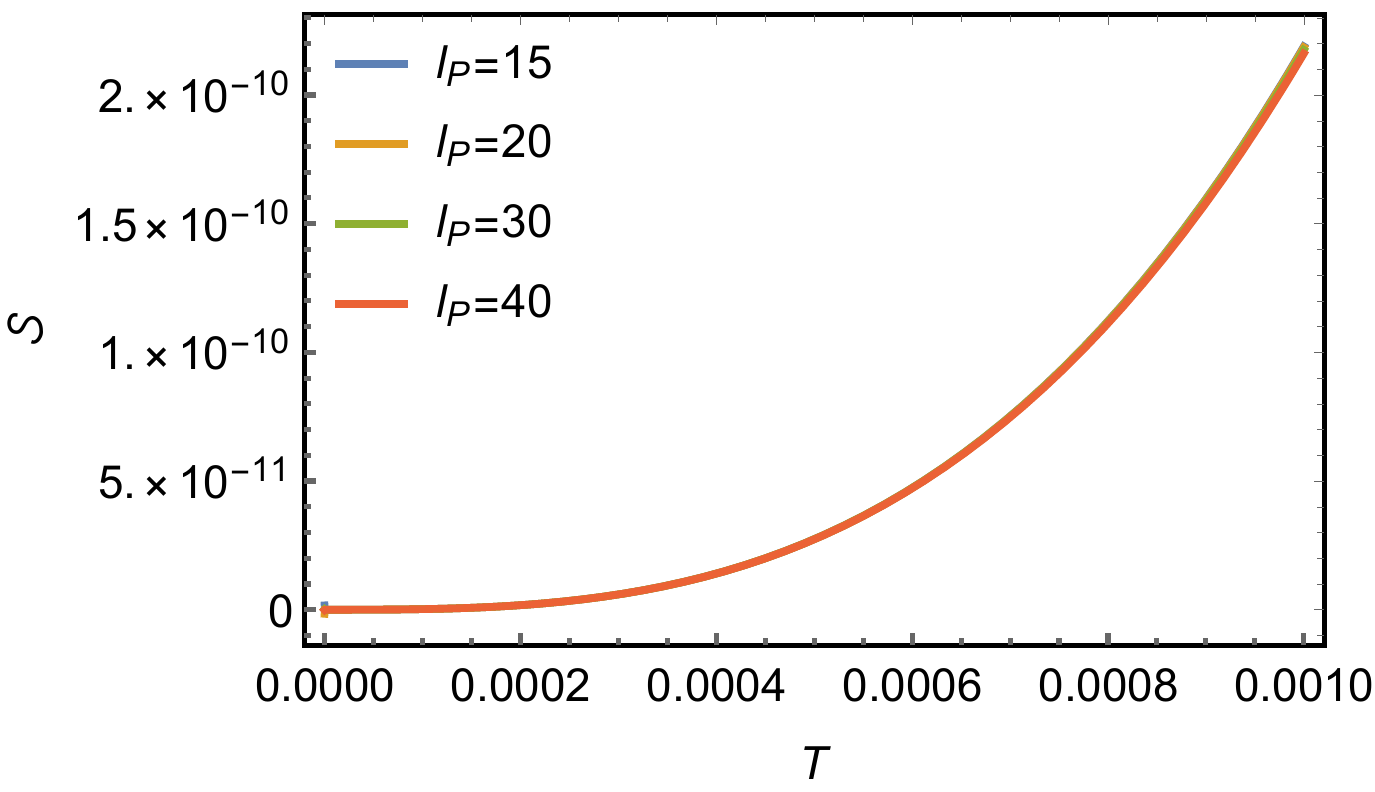}
\includegraphics[width=6cm,height=4cm]{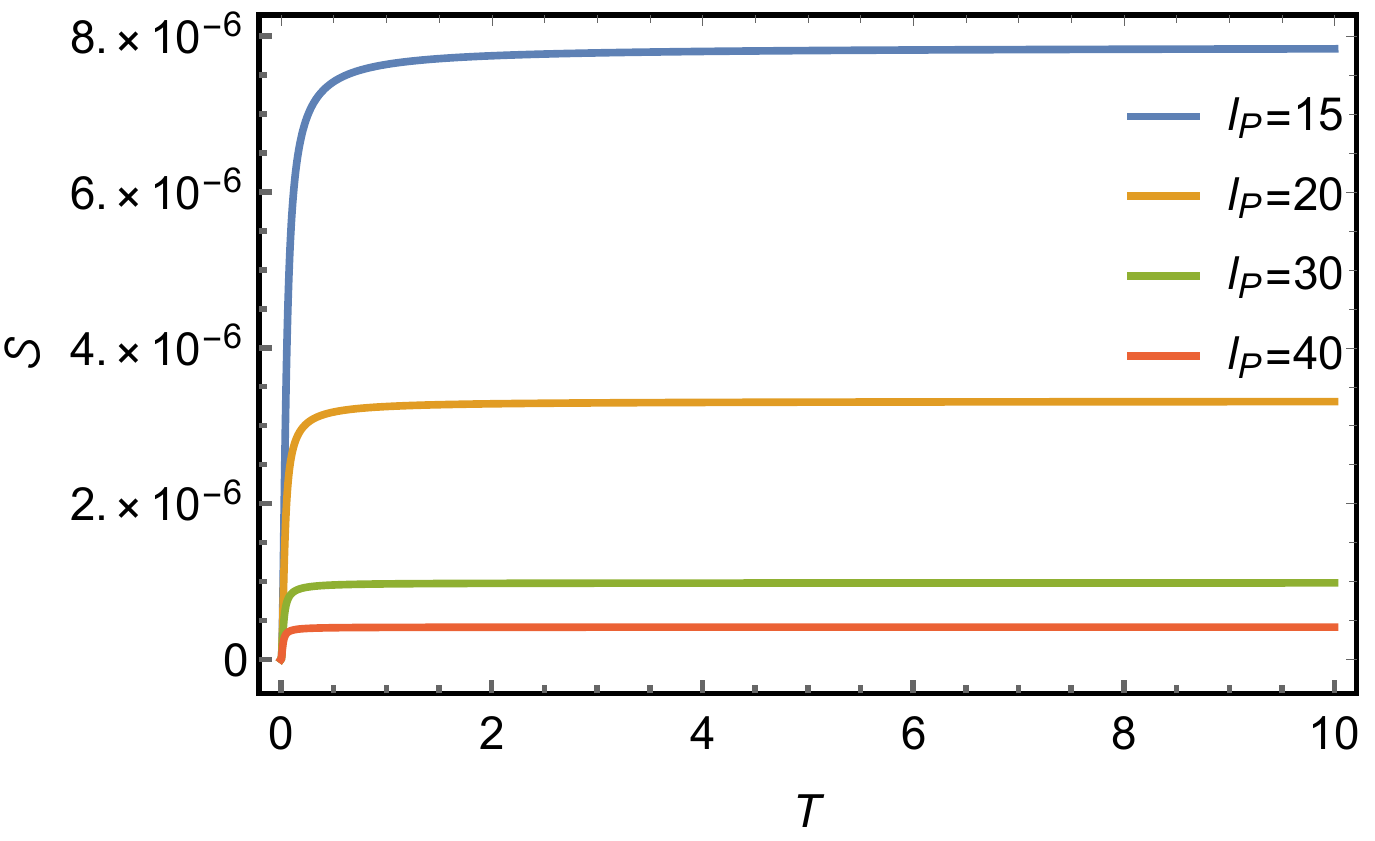}
\includegraphics[width=6cm,height=4cm]{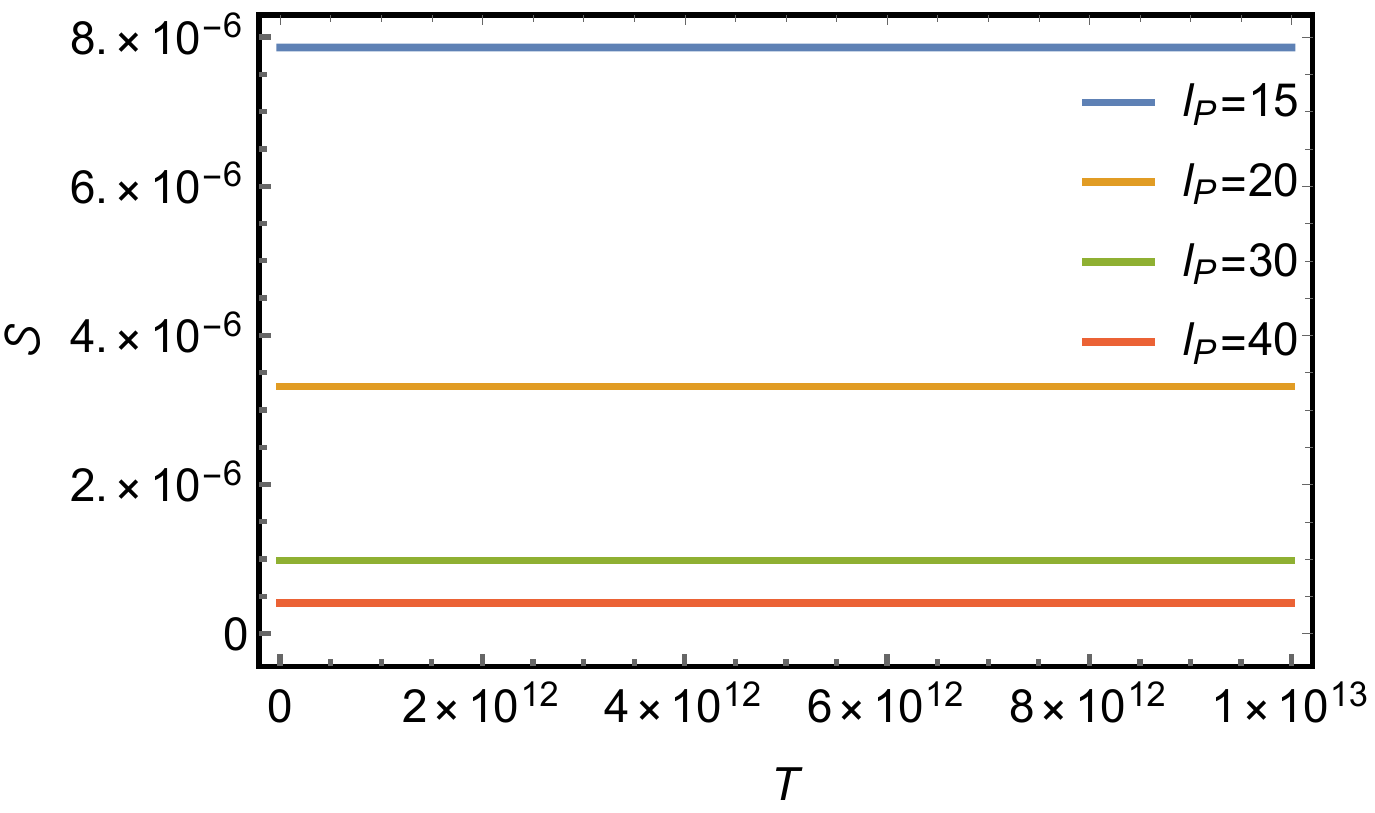}
\caption{Entropy for different values of $l_{P}$.}
\label{entropy}
\end{figure}

\begin{figure}[tbh]
\centering
\includegraphics[width=6cm,height=4cm]{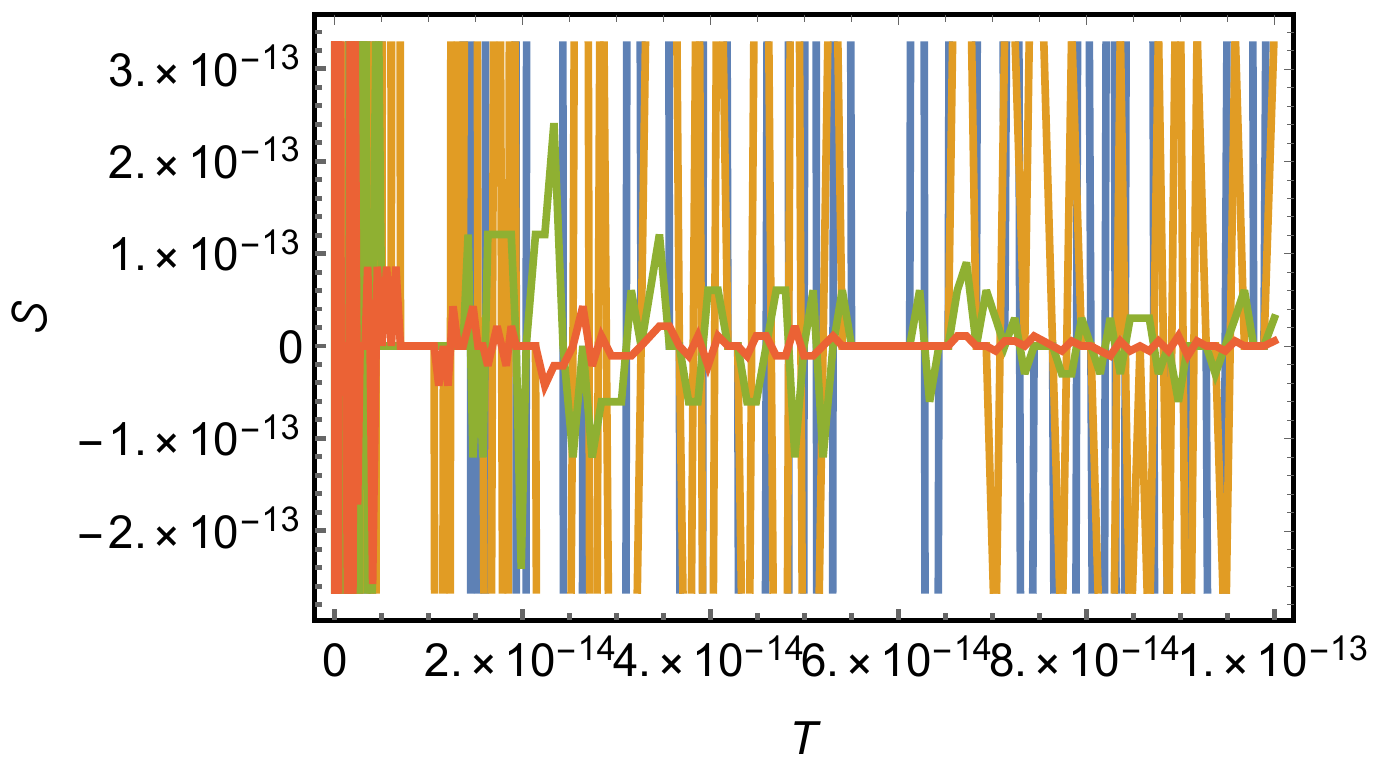}
\includegraphics[width=6cm,height=4cm]{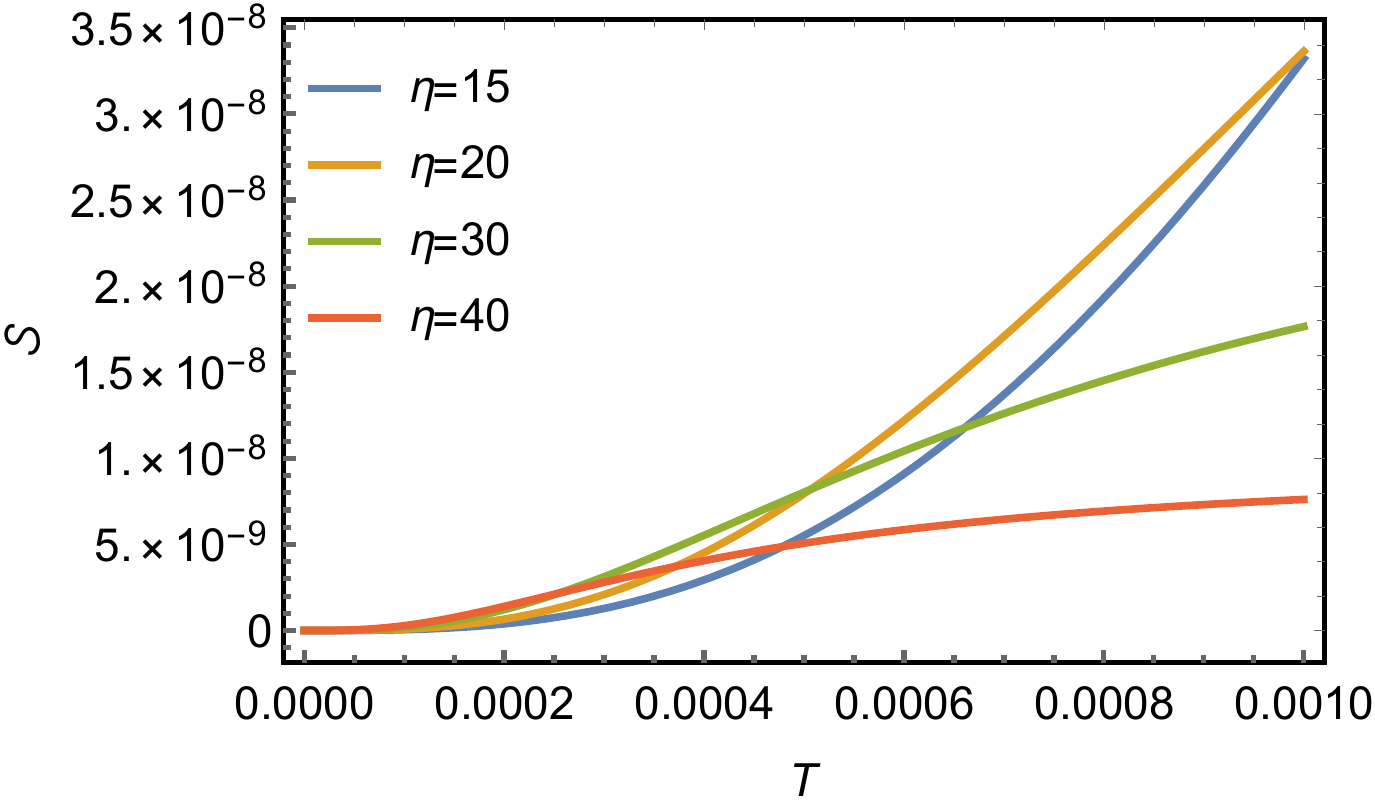}
\includegraphics[width=6cm,height=4cm]{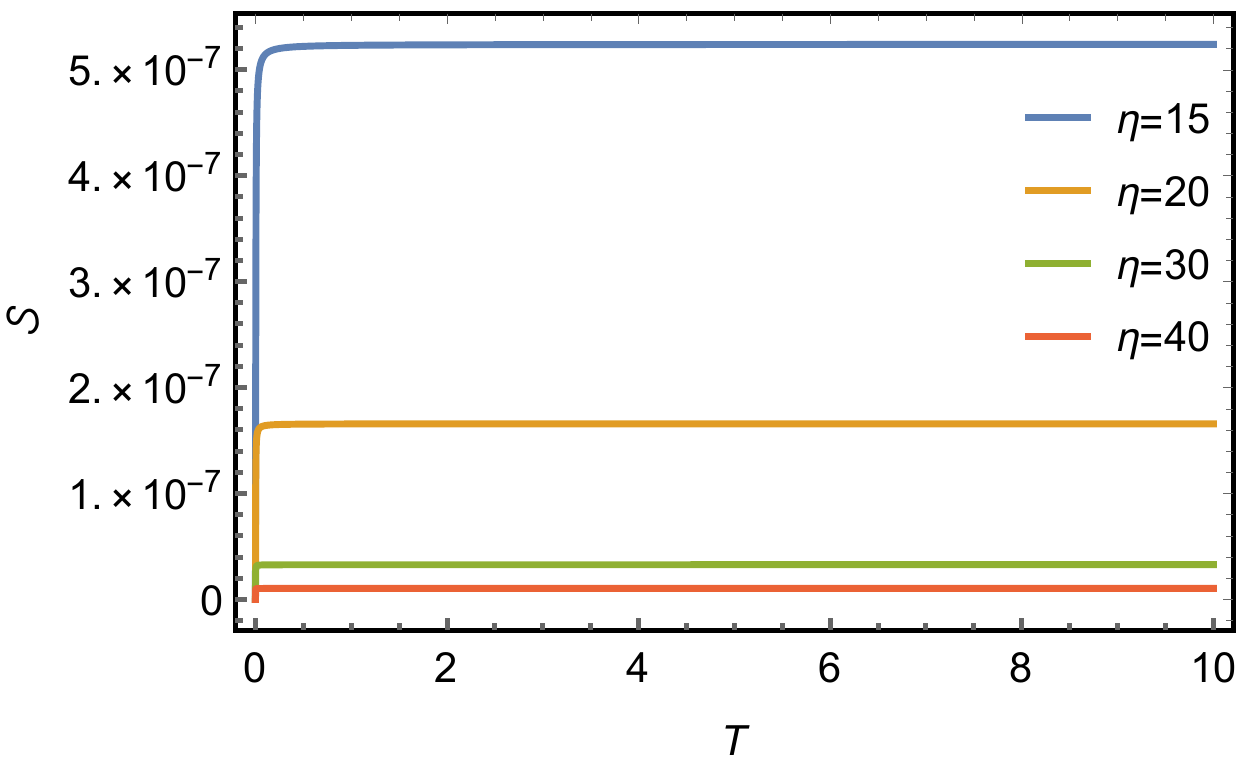}
\includegraphics[width=6cm,height=4cm]{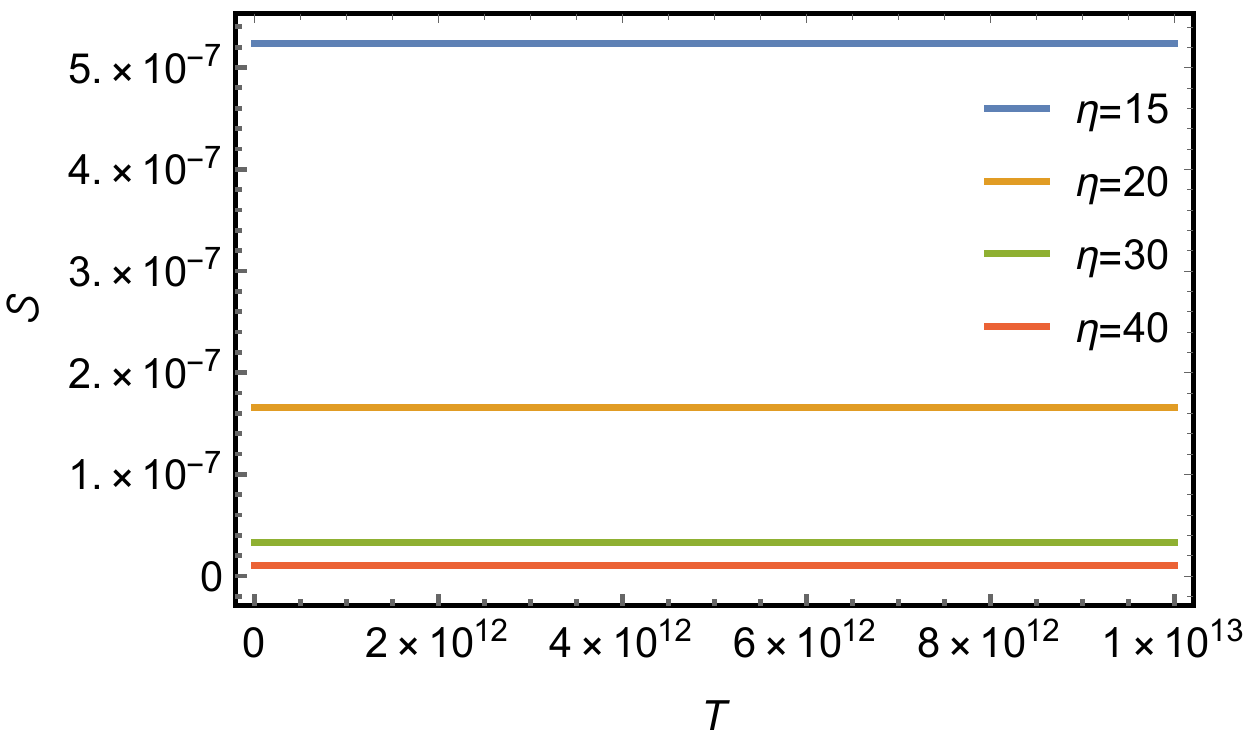}
\caption{Entropy for different values of $\eta$.}
\label{entropy2}
\end{figure}

The behavior of the entropy agrees with the previously known results from low through high-temperature regimes; in other words, this means that the second law of the thermodynamics is maintained, as it is natural to expect. Nevertheless, additionally of what was discussed in Refs. \cite{ling2009bouncing,pan2016bouncing}, a confusing behavior was also revealed in Figs. \ref{entropy} and \ref{entropy2} for the very low temperatures (the cosmic microwave background temperature): negative values of the entropy aroused. This might also be related to the bouncing region.

The critical temperature for both configuration of entropy and Helmholtz free energy was $T \leq 3.5 \times 10^{-5} $ GeV. In this regime, the system seems to display instability since the thermal quantities entropy and Helmholtz free energy showed fluctuation between negative and positive values.

Finally, the heat capacity is found to look like
\ie
\begin{split}
&C_{V}(\beta,\eta,l_{P})  =    \frac{\kappa_{B} \beta^{2}}{\pi^{2}} \int^{\infty}_{0} \frac{E^{2}}{\eta^{2} l^{2}_{P}} \sin^{2}(\eta^{2} l_{P} E)\cos(\eta^{2} l_{P} E) \frac{e^{-2\beta E}}{(1-e^{-\beta E})^{2}} \,\mathrm{d}E \\
& + \frac{\kappa_{B} \beta}{\pi^{2}} \int^{\infty}_{0} \frac{E^{2}}{\eta^{2} l^{2}_{P}} \sin^{2}(\eta^{2} l_{P} E)\cos(\eta^{2} l_{P} E) \frac{e^{-\beta E}}{1-e^{-\beta E}} 
\,\mathrm{d}E\\
= &  \frac{\beta}{72 \pi ^2 \eta ^2 l_{P}^2}\left\{ \frac{18 E^2 \cos \left(3 E \eta ^2 l_{P}\right)}{e^{\beta  E}-1} \right.\\
&\left. 9 \left[ -\frac{2 E^2 \cos \left(E \eta ^2 l_{P}\right)}{e^{\beta  E}-1} +\frac{1}{\eta ^4 l_{P}^2} e^{-i E \eta ^2 l_{P}}  \left[-l_{P} E \eta ^2  \left(E \eta ^2 l_{P}+2 i\right) \, _2F_1\left(1,-\frac{i l_{P} \eta ^2}{\beta };1-\frac{i l_{P} \eta ^2}{\beta };e^{\beta  E}\right) \right.\right.\right.\\
&\left.\left.\left.  +\left(-2+2 i E \eta ^2 l_{P}\right) \, _3F_2\left(1,-\frac{i l_{P} \eta ^2}{\beta },-\frac{i l_{P} \eta ^2}{\beta };1-\frac{i l_{P} \eta ^2}{\beta },1-\frac{i l_{P} \eta ^2}{\beta };e^{\beta  E}\right) \right.\right.\right.\\
&\left.\left.\left. +2 \, _4F_3\left(1,-\frac{i l_{P} \eta ^2}{\beta },-\frac{i l_{P} \eta ^2}{\beta },-\frac{i l_{P} \eta ^2}{\beta };1-\frac{i l_{P} \eta ^2}{\beta },1-\frac{i l_{P} \eta ^2}{\beta },1-\frac{i l_{P} \eta ^2}{\beta };e^{\beta  E}\right) \right] \right] \right. \\
&\left. - \frac{e^{i E \eta ^2 l_{P}}}{\eta ^4 l_{P}^2}  \left[ E \eta ^2 l_{P} \left(E \eta ^2 l_{P}-2 i\right) \, _2F_1\left(1,\frac{i l_{P} \eta ^2}{\beta };\frac{i l_{P} \eta ^2}{\beta }+1;e^{\beta  E}\right) \right.\right.\\
&\left.\left.+\left(2+2 i E \eta ^2 l_{P}\right) \, _3F_2\left(1,\frac{i l_{P} \eta ^2}{\beta },\frac{i l_{P} \eta ^2}{\beta };\frac{i l_{P} \eta ^2}{\beta }+1,\frac{i l_{P} \eta ^2}{\beta }+1;e^{\beta  E}\right) \right.\right.\\
& \left.\left.- 2 \, _4F_3\left(1,\frac{i l_{P} \eta ^2}{\beta },\frac{i l_{P} \eta ^2}{\beta },\frac{i l_{P} \eta ^2}{\beta };\frac{i l_{P} \eta ^2}{\beta }+1,\frac{i l_{P} \eta ^2}{\beta }+1,\frac{i l_{P} \eta ^2}{\beta }+1;e^{\beta  E}\right)\right] \right. \\
& \left. +\frac{e^{-3 i E \eta ^2 l_{P}}}{\eta ^4 l_{P}^2} \left[ 3 E \eta ^2 l_{P} \left(3 E \eta ^2 l_{P}+2 i\right) \, _2F_1\left(1,-\frac{3 i l_{P} \eta ^2}{\beta };1-\frac{3 i l_{P} \eta ^2}{\beta };e^{\beta  E}\right)   \right. \right.\\
& \left. \left.+\left(2-6 i E \eta ^2 l_{P}\right) \, _3F_2\left(1,-\frac{3 i l_{P} \eta ^2}{\beta },-\frac{3 i l_{P} \eta ^2}{\beta };1-\frac{3 i l_{P} \eta ^2}{\beta },1-\frac{3 i l_{P} \eta ^2}{\beta };e^{\beta  E}\right) \right. \right.\\
& \left. \left.+ 2 \, _4F_3\left(1,-\frac{3 i l_{P} \eta ^2}{\beta },-\frac{3 i l_{P} \eta ^2}{\beta },-\frac{3 i l_{P} \eta ^2}{\beta };1-\frac{3 i l_{P} \eta ^2}{\beta },1-\frac{3 i l_{P} \eta ^2}{\beta },1-\frac{3 i l_{P} \eta ^2}{\beta };e^{\beta  E}\right)    \right] \right. \\
& \left. +\frac{e^{3 i E \eta ^2 l_{P}}}{\eta ^4 l_{P}^2} \left[ 3 E \eta ^2 l_{P} \left(3 E \eta ^2 l_{P}-2 i\right) \, _2F_1\left(1,\frac{3 i l_{P} \eta ^2}{\beta };1+\frac{3 i l_{P} \eta ^2}{\beta };e^{\beta  E}\right)   \right. \right.\\
& \left. \left.+\left(2+6 i E \eta ^2 l_{P}\right) \, _3F_2\left(1,\frac{3 i l_{P} \eta ^2}{\beta },+\frac{3 i l_{P} \eta ^2}{\beta };1+\frac{3 i l_{P} \eta ^2}{\beta },1+\frac{3 i l_{P} \eta ^2}{\beta };e^{\beta  E}\right) \right. \right.\\
& \left. \left.+ 2 \, _4F_3\left(1,\frac{3 i l_{P} \eta ^2}{\beta },\frac{3 i l_{P} \eta ^2}{\beta },\frac{3 i l_{P} \eta ^2}{\beta };1+\frac{3 i l_{P} \eta ^2}{\beta },1+\frac{3 i l_{P} \eta ^2}{\beta },1+\frac{3 i l_{P} \eta ^2}{\beta };e^{\beta  E}\right)    \right]
\right\} \Bigg\rvert^{\infty}_{0} = 0.
\end{split}
\fe
Therefore, just as occurred for the mean energy, the heat capacity turned out to yield only the trivial contribution after performing the limits of integration.  

%%%%%%%%%%%%%%%%%%%%%%%%%%%%%%%%%%%%%%%%%%%%%%%%%%%%%%%%%%%%%%%%%%%%%%%%%%%%%%%%%%%%%%%%%%%%%%%%%%%%%%%%%%%%%%%%%%%%%%%%%%%%%%%%%%%%%%%%%%%%%%%%%%%%%%%%%%%%%%%%%%%%%%%%%%%%%%%%%%%%%%%%%%%%%%%%%%%%%%%%%%%%%%%%%%%%%%%%%%%%%%%%%%%%%%%%%%%%%%%%%%%%%%%%%%%%%%%%%%%%%%%%%%%%%%%%%%%%%%%%%%%%%%%%%%%%%%%%%%%%%%%%%%%%%%%%%%%%%%%%%%%%%%%%%%%%%%%%%%%%%%%%%%%%%%%%%%%%%%%%%%%%%%%%%%%%%%%%%%%%%%%%%%%%%%%%%%%%%%%%%%%%%%%%%%%%%%%%%%%%
\section{Conclusion} \label{conclusion}

In this work, we focused on examining the thermal behavior of a photon gas within the context of bouncing universe. To accomplish this, we started from a modified dispersion relation which accounted for modified Friedmann equations with a bounce solution. All of our results were derived \textit{analytically}. Furthermore, we considered three different scenarios of temperature of the universe: inflationary epoch, electroweak era and cosmic microwave background. Initially, we calculated the accessible states of the system. With it, we performed an analysis of the impact of the parameters $l_{P}$ and $\eta$ on the modification of the thermodynamic properties of interest.

The spectral radiance $\chi(\nu)$ turned out to have an intriguing behavior, i.e., it had a fluctuation between positive and negative frequencies. Perhaps, it could be a consequence of either an disturbing issue due to the bouncing solution or for the periodic aspect of the trigonometric function which came from the dispersion relation.

More so, we calculated the mean energy, the entropy, the Helmholtz free energy, and the heat capacity as well. Nevertheless, one intriguing aspect also brought out: after applying the integration limits, the mean energy and the heat capacity turned out to have no contribution in our calculations. On the other hand, the Helmholtz free energy and the entropy demonstrated to be consistent with previous studies, i.e., the second law of the thermodynamics was maintained. However, when we considered the low temperature regime, such thermal functions exhibited a fluctuation aspect as appeared in the spectral radiance. The critical temperature for both configurations of entropy and Helmholtz free energy were $T \leq 3.5 \times 10^{-5} $ GeV. In this regime, the system seemed to have instability since such quantities showed fluctuation between negative and positive values. Such a phenomenon might have happened due to proximity to the bouncing point. Nevertheless, further investigation might be accomplished in order answer properly this issue.

%%%%%%%%%%%%%%%%%%%%%%%%%%%%%%%%%%%%%%%%%%%%%%%%%%%%%%%%%%%%%%%%%%%%%%%%%%%%%%%%%%%%%%%%%%%%%%%%%%%%%%%%%%%%%%%%%%%%%%%%%%%%%%%%%%%%%%%%%%%%%%%%%%%%%%%%%%%%%%%%%%%%%%%%%%%%%%%%%%%%%%%%%%%%%%%%%%%%%%%%%%%%%%%%%%%%%%%%%%%%%%%%%%%%%%%%%%%%%%%%%%%%%%%%%%%%%%%%%%%%%%%%%%%%%%%%%%%%%%%%%%%%%%%%%%%%%%%%%%%%%%%%%%%%%%%%%%%%%%%%%%%%%%%%%%%%%%%%%%%%%%%%%%%%%%%%%%%%%%%%%%%%%%%%%%%%%%%%%%%%%%%%%%%%%%%%%%%%%%%%%%%%%%%%%%%%%%%%%%%%

\section*{Acknowledgments}
\hspace{0.5cm} The work by A. Yu. Petrov. has been partially supported by the CNPq project No. 301562/2019-9. A. A. Araújo Filho acknowledges the Facultad de Física - Universitat de València and Gonzalo J. Olmo for the kind hospitality when this work was made. Moreover, A. A. Araújo Filho has been partially supported by Conselho Nacional de Desenvolvimento Cient\'{\i}fico e Tecnol\'{o}gico (CNPq) - 142412/2018-0, and CAPES-PRINT (PRINT - PROGRAMA INSTITUCIONAL DE INTERNACIONALIZAÇÃO) - 88887.508184/2020-00. Most of the calculations presented in this manuscript were accomplished by using the \textit{Mathematica} software. More so, the authors would like to thank J. A. A. S. Reis for the help with the computer calculations.

\bibliographystyle{ieeetr}
\bibliography{main}

\end{document}